\documentclass[prb,twocolumn,superscriptaddress,showpacs]{revtex4}
\usepackage{amsmath}
\usepackage{epsfig}
\usepackage{dcolumn}% Align table columns on decimal point
\usepackage{bm}% bold math
\usepackage{slashed}
\usepackage[toc,page]{appendix}
\usepackage{float}
\usepackage{float}
\usepackage{amsmath}
\usepackage{amssymb}
\usepackage{epsfig}

\usepackage{graphicx}

\def\gapp{\lower.35em\hbox{$\stackrel{\textstyle>}{\sim}$}}
\def\lapp{\lower.35em\hbox{$\stackrel{\textstyle<}{\sim}$}}

\begin{document}
\bibliographystyle{apsrev}
%

%\draft
\title{Effect of finite temperature and uniaxial anisotropy on the Casimir effect with three-dimensional topological insulators}

\author{Adolfo G. Grushin}
\affiliation{Instituto de Ciencia de Materiales de Madrid, CSIC, Cantoblanco, E-28049 Madrid, Spain.}

\author{Pablo Rodriguez-Lopez}
\affiliation{Dept. de F\'isica Aplicada I and GISC, Universidad Complutense, 28040 Madrid, Spain}

\author{Alberto Cortijo}
\affiliation{Departamento de F\'{i}sica te\'{o}rica, Universidad Aut\'{o}noma de Madrid, E-28049, Madrid, Spain}

\date{\today}
%%%%%%%%%%%%%%%%%%%%%%%%%%%%%%%%%%%%%%%%%%%%%%%%%%%%%%%%%%%%%%%%%%%%%%%%%%%%%
\begin{abstract}
In this work we study the Casimir effect with three-dimensional topological insulators including the effects of temperature and uniaxial anisotropy. Although precise experimental values for the optical properties of these materials are yet to be established, qualitative analysis is still possible. We find qualitatively that the reported repulsive behavior and the equilibrium point are robust features of the system, and are favored by low temperatures and the enhancement of the optical response parallel to the optical axis. The dependence of the equilibrium point with temperature and with the topological magnetoelectric polarizability characteristic of three-dimensional topological insulators is also discussed.
\end{abstract}

\maketitle

\section{Introduction}
Since H.B.G. Casimir wrote his seminal paper \cite{Cas48} proposing that two uncharged metallic plates should attract in vacuum due to quantum fluctuations still little is known of how to revert the sign of the force and obtain repulsion or levitation. The generalization of the theory to include dielectric bodies \cite{DLP61} has made it possible to achieve, very recently, repulsion with test bodies immersed in a dielectric fluid \cite{MCP09}. However, achieving repulsion in vacuum, which would have important consequences for device applications \cite{KMM09,BKMM09}, is still to be experimentally tested. Theoretically, there have been several proposals to achieve such situation. The first proposal considered magnetic versus non-magnetic situations \cite{B74} and more recent proposals include the use of metamaterials \cite{RDM06,LP07,ZZK09a} and geometry to induce repulsive scenarios \cite{LMR10}. Importantly, it has also been shown that for general dielectric-dielectric situations, there is no repulsive behavior in vacuum \cite{RK09}. In a recent paper \cite{GC10}, two of us suggested that repulsive behavior could be achieved by using topological insulators \cite{Moore10,HQW08,HXW09,HK10}. The topological nature of these materials \cite{QHZ08,EMB09} provides a magnetoelectric term modifying the response to an external electromagnetic field, which can then cause the Casimir effect to reverse sign at short distances and remain attractive at large distances\cite{GC10}. Consequently, at an intermediate distance $d_{eq}$, the Casimir force vanishes and a stable configuration is possible.\\
In this work, we aim to include the effect of temperature and uniaxial anisotropy to this behavior, two relevant factors in experimental situations. It is well known by know that anisotropy can modify the Casimir force by changing the form of the reflection coefficients \cite{BKMM09}, and has been studied in the context of repulsive interactions \cite{DLL08}, even including the effect of temperature \cite{RDM08}. In the present paper we address the question of whether uniaxial anisotropy and temperature can enhance or destroy the repulsive behavior in the context of topological insulators, and in what cases can this occur. To achieve this goal we derive the reflection coefficients for a generic anisotropic topological insulator, a result which to our knowledge is absent in the topological insulator literature. With the help of these, we find that low temperatures favor the repulsive behavior and we determine that enhancing uniaxial anisotropy in the direction parallel to the optical axis also acts to increase repulsion.\\
To our knowledge, precise optical characterization of these materials is still lacking, and thus it remains an open question if real materials are to show repulsion at these distances. Nevertheless, throughout the paper, we use choices of parameters which can be relevant for topological insulators such as TlBiSe$_2$\cite{Chen10} (see also concluding paragraph in Ref.\cite{GC10}) which are mainly low dielectric permeability at zero frequency $\epsilon(0)$ and small magnetoelectric polarizability $\theta$. We do not restrict the study only to these values but explore other regions of parameters for completeness.\\
The work is structured as follows. Section \ref{sec:T=0} is devoted to review the formalism and notation on which we rely on. To do so, we will briefly review the isotropic case at zero temperature ($T=0$) discussed in Ref. \cite{GC10}, to which we will add the dependence of the minimum with the topological magnetoelectric polarizability, not discussed in Ref. \cite{GC10}. Details of the proof of the existence of repulsion are left for appendix \ref{App:Minimum}. In section \ref{sec:Tneq0} we will study the effect of temperature on the Casimir energy as a function of distance, both analytically and numerically. We will also discuss the position of the minimum as a function of temperature and topological magnetoelectric polarizability $\theta$. Details of the analytical results can be found in appendix \ref{App:class}. In section \ref{sec:uniaxial} we will discuss the effect of uniaxial anisotropy on the Casimir energy at $T=0$. First, we will explicitly write the Fresnel coefficients for this case, leaving the details of the derivation for appendix \ref{App:coef}. In the remaining part of this section we will consider different values for the relevant parameters and discuss their effect on the position of the minimum. Finally, in section \ref{sec:uniaxialandtemp} we will present the results for the case of uniaxially anisotropic topological insulator at $T\neq0$ to end with section \ref{sec:conc}, which includes a summary of results and some concluding remarks regarding real systems.\\
We emphasize that the numerical results presented in this paper explore regions of parameters without any restricting assumption other than positive energy condition inherent to magnetoelectric couplings, discussed in section \ref{sec:Tneq0}.B. On the other hand, in order to derive analytical results, some feasible assumptions are to be considered, such as small magnetoelectric coupling, all of which apply only to the results in appendix \ref{App:class}, and hence will be discussed in the mentioned appendix.

\section{\label{sec:T=0}Casimir-Lifshitz force between topological insulators at $T=0$}

It is well known from different approaches, that the Casimir energy density stored by two dielectric parallel plates (labeled $1$ and $2$) of area $A$, separated by a distance $d$, at zero temperature ($T=0$) can be written as\cite{DLP61,Reyn91,RE09,BKMM09}:
\begin{equation}
\label{CasimirEnergy}
\frac{E_{c}(d)}{A\hbar} = \int_0^{\infty} \hspace{-1pt} \frac{d\xi}{2\pi} \int \frac{d^2 {\bf k}_{\|}}{(2\pi)^2} \log \det \left[1 - {\bf R}_1 \cdot {\bf R}_2 e^{-2 k_3 d}\right] ,
\end{equation}
where $k_3=\sqrt{\bm{k}^{2}_{\|}+ \xi^2/c^2}$ is the wave vector perpendicular to the plates, $\bm{k}_{\|}$ is the vector parallel to the plates and $\xi$ is the imaginary frequency defined as $\omega = i\xi$. The matrices ${\bf R}_{1,2}$  are $2\times2$ reflection matrices containing the Fresnel coefficients such that:
\begin{eqnarray}
\label{eq:ReflectionMatrices}
{\bf R} = \left[
\begin{array}{cc}
   R_{s,s} (i \xi, {\bf k}_{\|}) &  R_{s,p} (i \xi, {\bf k}_{\|}) \\
  R_{p,s} (i\xi, {\bf k}_{\|}) &  R_{p,p} (i \xi, {\bf k}_{\|})
\end{array} \right] ,
\end{eqnarray}
where $R_{i,j}$ describes the reflection amplitude of an incident wave with polarization $i$ which is reflected with polarization $j$. The label $s$ ($p$), equivalent to $TE$ ($TM$) modes, describes parallel (perpendicular) polarization of the electric field with respect to the plane of incidence. The Casimir force per unit area on the plates is obtained by differentiating expression (\ref{CasimirEnergy}): $F=-\partial_{d}E_{c}(d)$. A negative (positive) force, or equivalently a positive (negative) slope of $E_{c}(d)$, corresponds to attraction (repulsion) of the plates.\\
In order to calculate the Casimir energy one needs to compute the reflection coefficients. These coefficients relate the amplitude of the incident and reflected electric fields for different polarizations. Since the normal components of $\bm{D}$ and $\bm{B}$ and tangential components of $\bm{E}$ and $\bm{H}$ must be continuous along the interface, one can obtain the reflection amplitudes by solving Maxwell's equations at each side and then imposing the mentioned matching conditions for the fields (see below).

\begin{figure}
\includegraphics[scale=0.2]{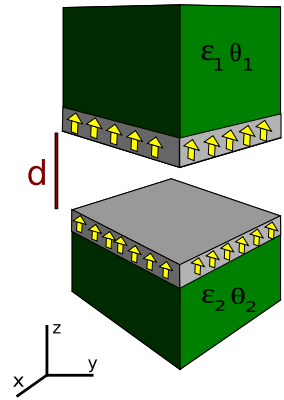}
\caption{\label{Fig:plates}(color online)Topological insulating plates separated by a distance $d$ covered with a thin ferromagnetic layer of thickness $l<<d$ (not to scale). Changing the sign of the magnetization of one of the plates (parallel or antiparallel to the surface normal) is equivalent to changing the sign of (either) $\theta_{1,2}$.}
\end{figure}

\subsection{Topological insulators: The model}

Following Ref.\cite{GC10} we will consider a Casimir system composed by two topological insulator plates separated by a distance $d$, shown in Fig.\ref{Fig:plates}. The electromagnetic response of three dimensional topological insulators, which determines the reflection coefficients, is governed by the lagrangean:
\begin{eqnarray}\label{eq:lagrangean}
\mathcal{L} = \mathcal{L}_{0} + \mathcal{L}_{\theta} = \left(\bm{E}\cdot\bm{D}+\bm{B}\cdot\bm{H}\right)
\end{eqnarray}
with the difference that now the constitutive relations for the electric displacement field $\bf{D}$ and the magnetic field $\bf{H}$ are given by:

\begin{eqnarray}\label{eq:D}
D_{i} &=& \epsilon_{ij}E_{j}+\dfrac{\alpha\theta}{\pi}B_{i}\\
\label{eq:H}
H_{i} &=& (\mu)^{-1}_{ij}B_{j}-\dfrac{\alpha\theta}{\pi}E_{i}.
\end{eqnarray}

The first term of each constitutive equation is the usual electromagnetic term defined in terms of the dielectric function $\epsilon(\omega)$ and magnetic susceptibility $\mu(\omega)$, giving rise to the ordinary electromagnetic term in the lagrangean \eqref{eq:lagrangean} which we label $\mathcal{L}_0$. We restrict ourselves to non magnetic topological insulators so $\mu=1$.  The second term in each constitutive equation is a non trivial axionic \cite{Wilczek87} or topological magnetoelectric term \cite{QHZ08,EMB09} which gives rise to a topological $\mathcal{L}_{\theta}$ term in the lagrangean. Time reversal symmetry indicates that $\theta=0,\pi$ (mod $2\pi$) being $\theta=\pi$ the case for topological insulators and $\theta=0$ the case for trivial insulators.\\
When the boundary is included, the action corresponding to the lagrangean \eqref{eq:lagrangean}, $S_0+S_{\theta}$, is a fair description of the topological insulator only when a time reversal breaking perturbation is induced on the surface to gap the surface states\cite{QHZ08}. Along the lines of similar situations \cite{QHZ08,QLZZ09,GC10}, in this work we consider that the time reversal perturbation is a magnetic coating of small thickness $l<<d$ ($d$ will be of the order of $\mu$m) which gaps the surface states. In the described situation, $\theta$ is quantized in odd integer values of $\pi$ such that:

\begin{eqnarray}
\theta = (2n + 1)\pi ,
\end{eqnarray}

where $n \in \mathbb{Z}$, determined by the nature of the coating, but independent of the absolute value of the magnetization of the coating. Positive or negative values of $\theta$ are related to different signs of the magnetization on the surface \cite{QLZZ09}, which we consider is perpendicular to the plane of the plates. Each topological insulator plate is characterized by the frequency dependent dielectric function $\epsilon(\omega)$ and the quantized magnetoelectric term $\theta$. Being a topological contribution, $\theta$ is defined in the bulk as a constant whenever the bulk Brillouin zone is defined \cite{EMB09}. Nevertheless, at the surface, the effective action must be valid as long as the condition $l<<d$ is fulfilled, where magnetic effects from the covering magnetic layer could renormalize its value, and the system should be treated as a layered system. We henceforth assume that this is not the case and consider only the region where $l<<d$ (nm $<< \mu$m in practical terms). \\
The parasitic magnetic forces between the magnetic layers can be estimated following Ref.\cite{Bruno02}. The dipole-dipole interaction is of the order of attoN at distances of 50 nm and the magnetic Casimir force\cite{Bruno02} is $\sim 1 fN$. Since we will be dealing with distances of the order of $\mu$m the magnetic forces will be even smaller compared to the Casimir force that is of the order of $p$N\cite{GC10}. Thus, we neglect the parasitic magnetic interactions, and focus only on the Casimir force experienced by the plates.

\subsection{Results for isotropic topological insulator plates at $T=0$}

In this section we review for completeness the case of two isotropic topological insulator plates discussed in Ref.\cite{GC10} and extend some of the results. The topological part $S_{\theta}$ in the action does not modify Maxwell's equations as long as the constituent relations are changed according to \ref{eq:D} and \ref{eq:H}, taken at this stage to be isotropic.\\
From these, we can obtain the reflection matrices \eqref{eq:ReflectionMatrices} by imposing the continuity of the tangential component of $\bf{H}$ and the normal component of $\bf{D}$. For a plate described by the optical responses $\theta_{i}$ and $\epsilon_{i}(\omega)$ immersed in vacuum the reflection matrices are given by \cite{Obu09,CY09}:

\begin{widetext}
\begin{equation}
%\eqname{S.1}
\label{eq:ReflectionMatricesTIcomplete}
{\bf R}_i = \dfrac{1}{\Delta}\left(
\begin{array}{cc}
  1-n^2_{i}-\bar{\alpha}_{i}^2 + n_{i}\chi_{-} &  2 \bar{\alpha}_{i} \\
  2\bar{\alpha}_{i} & -1+n^2_{i}+\bar{\alpha}_{i}^2 + n_{i}\chi_{-}
\end{array} \right),
\end{equation}
\end{widetext}

where $i=1,2$ labels each plate, $\bar{\alpha}_{i}=\frac{\alpha\theta_{i}}{\pi}$, $\alpha$ is the fine structure constant ($\alpha = \frac{e^2}{\hbar c}$), $n_i = \sqrt{\epsilon_{i}(\omega)}$ is the refractive index of each plate, $\Delta = 1+n^2_{i}+\bar{\alpha}_{i}^2 + n_{i}\chi_{+}$ and:

\begin{eqnarray}
\label{eq:Xipm}
 \chi_{\pm}= %\dfrac{\cos\beta_{i}}{\cos\beta_{t}}\pm\dfrac{\cos\beta_{t}}{\cos\beta_{i}},
 \dfrac{\xi^2+\mathbf{k}_{\|}^2 \pm \left(\xi^2 + \frac{\mathbf{k}_{\|}^2}{n_{i}^2}\right)}{\sqrt{\left(\xi^2 + \mathbf{k}_{\|}^2\right)\left(\xi^2 + \frac{\mathbf{k}_{\|}^2}{n_{i}^2}\right)}}.
\end{eqnarray}

It is possible to show that when $\bar{\alpha}=0$ and after a little algebra, these coefficients reduce to the ordinary Fresnel coefficients for a dielectric-dielectric interface \cite{Born99}. By its definition, $\Delta$ is always positive and the off-diagonal terms have a an overall sign governed by the sign of $\theta_{i}$. \\
Introducing these matrices in \eqref{CasimirEnergy} and assuming that both topological insulators are described by an oscillator model of the form:

\begin{equation}\label{eq:phendiel}
\epsilon(i\xi)=1+\sum_{i}\dfrac{\omega_{e,i}^2}{\xi^2+\omega^2_{R,i}+\gamma_{R,i}\xi},
\end{equation}

it was shown\cite{GC10} that two different situations could arise: 1) when the topological terms had the same signs, i.e. $\mathrm{sgn}(\theta_{1})=\mathrm{sgn}(\theta_{2})$ the Casimir force was attractive for all distances, and 2) when the signs were opposite $\mathrm{sgn}(\theta_{1})=-\mathrm{sgn}(\theta_{2})$, it was shown both analytically and numerically that a repulsive region at small distances appeared. At large distances, attraction was recovered and the two regimes were separated by a minimum in the Casimir energy; an equilibrium point. The dependence of the minimum was studied for a one oscillator model, showing that the repulsive behavior was favored by reducing the ratio $\frac{\omega_{e}}{\omega_{R}}$. A detailed version of the proof of the existence of a minimum, together with the analytic analysis of the large and short distances limits of the Casimir energy are given in appendix \ref{App:Minimum}.

\begin{figure*}
\begin{minipage}{.49\linewidth}
\includegraphics[scale=0.47]{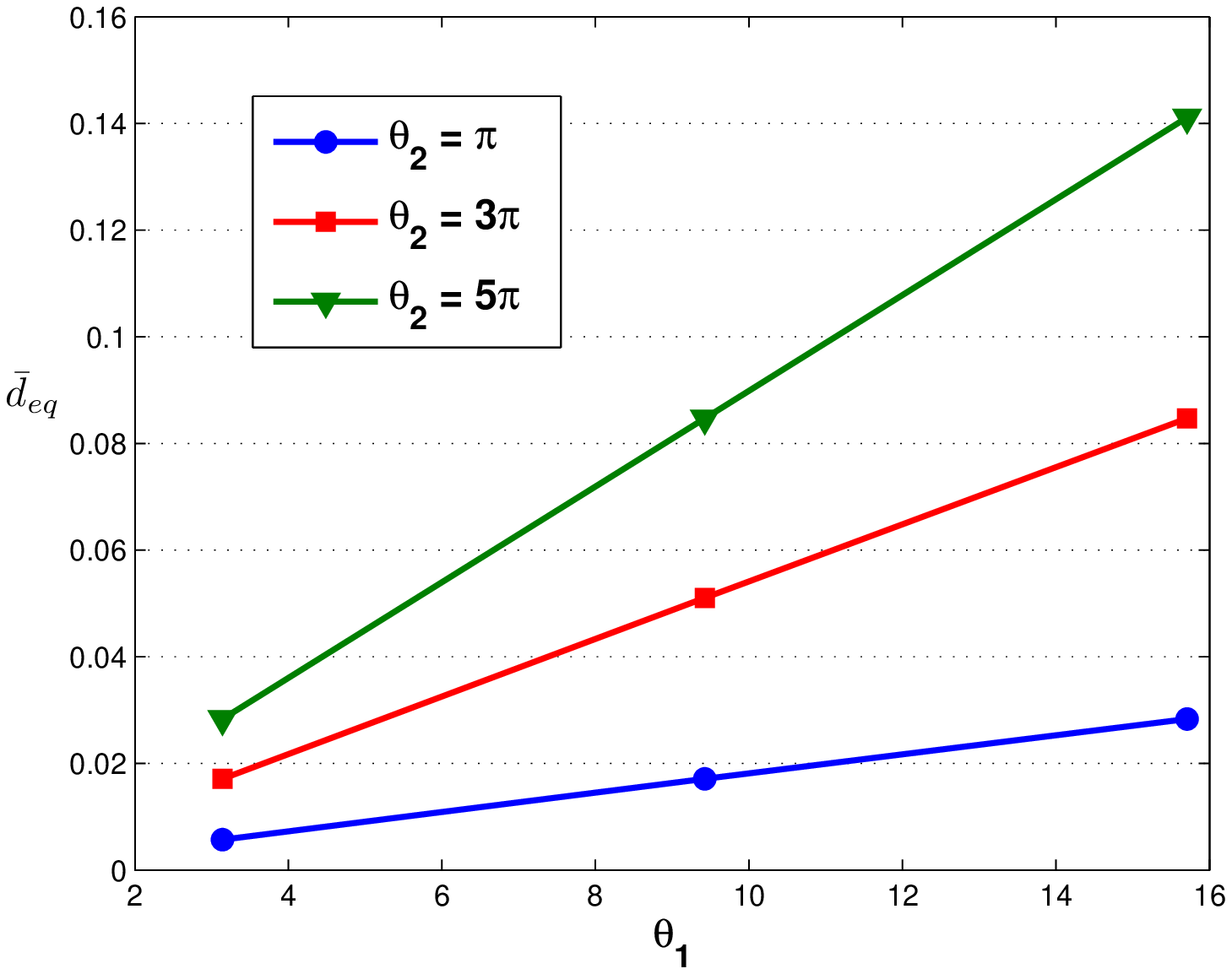}
(a)
\end{minipage}
\begin{minipage}{.49\linewidth}
\includegraphics[scale=0.47]{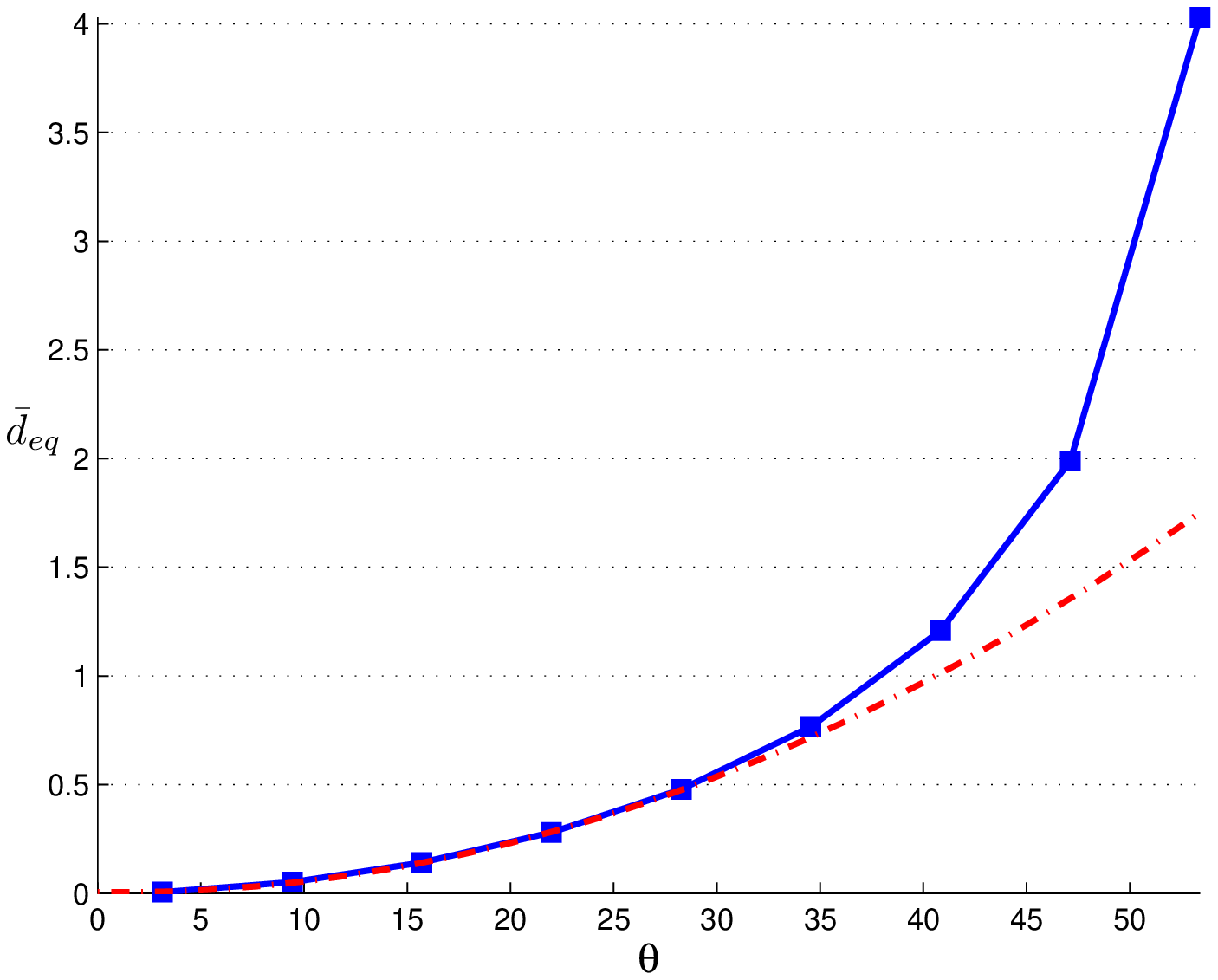}
(b)
\end{minipage}
\caption{\label{deqvstheta}(color online)(a) $\bar{d}_{eq}$ as a function of $\theta_{1}$ for different values of $\theta_{2}$ ranging from $\pi$ to $5\pi$. The solid blue line represents $\bar{d}_{eq}\propto |\theta_{1}\theta_{2}|$.(b) $\bar{d}_{eq}$ as a function of $\theta \equiv \theta_{1}=-\theta_{2}$ (black curve with dots). The behavior deviates from $\theta^2$ (bottom curve red) reasonable for small $\theta$. Beyond $\theta=17\pi$ the equilibrium point is lost (see text). For both figures the remaining parameters of the model where fixed to $\frac{\omega_{e}}{\omega_{R}}=0.45$}
\end{figure*}

To end this section we complete the analysis made in Ref.\cite{GC10} by studying the explicit dependence of the position of the equilibrium point ($d_{eq}$) with $\theta_{1}$ and $\theta_{2}$. We define for convenience the adimensional distance $\bar{d} \equiv \frac{\omega_{R}}{c}d$. Whenever the condition $\mathrm{sgn}(\theta_{1})=-\mathrm{sgn}(\theta_{2})$ is satisfied, it is easy to study numerically the dependence of $\bar{d}_{\mathrm{eq}}$, the equilibrium position as a function of both $\theta_{1,2}$. The analysis is shown in Fig.\ref{deqvstheta}. In Fig.\ref{deqvstheta}(a) we plot $\bar{d}_{\mathrm{eq}}$, against $\theta_{1}$ for different values of $\theta_{2}$ ranging from $\pi$ to $5\pi$. A numerical fit indicates that for small magnetoelectric couplings the dependence is given by:

\begin{eqnarray}\label{deqproptheta}
\bar{d}_{eq}\propto |\theta_{1}\theta_{2}|.
\end{eqnarray}

For higher magnetoelectric couplings, the behavior deviates from the simple $\theta^2$ law. This is shown in Fig.\ref{deqvstheta}(b) where we plot $\theta\equiv \theta_{1}=-\theta_{2}$. For high values of $\theta$ the equilibrium position grows faster to larger distances. Beyond a critical value, the magnetoelectric part of the reflection coefficients overwhelms the ordinary dielectric part giving rise to purely repulsive behavior and the equilibrium position disappears which is analogous to the $\epsilon(0)\to 1$ limit discussed in Ref. \cite{GC10}. The inclusion of temperature will modify further \eqref{deqproptheta}, as shown in the next section.\\
As a final comment for this section we mention that the obtained dependence of the minimum with $\theta$ suggests that an analytic way of deriving this result might be possible, especially in the case where $\theta_{1}=-\theta_{2}$. Nevertheless, the derivation of this result is not transparent from this formalism, although other formalisms\cite{RE09} can shed more light to the issue.

\section{\label{sec:Tneq0}Casimir-Lifshitz force between isotropic topological insulators at $T\neq0$}

In this section we include the effect of temperature for isotropic plates and we discuss how the repulsive behavior can be affected by changes in temperature.

\subsection{Inclusion of finite temperature effects}

To take into account finite temperature effects within the Lifshitz theory one must assume that the system is in thermal equilibrium. As a first step, the formal replacement \cite{BKMM09,RE09}:

\begin{eqnarray}\label{eq:subs}
\dfrac{\hbar}{2\pi}\int^{\infty}_{0}d\xi \longleftrightarrow k_{B}T{\sum^{\infty}_{l=0}}' ,
\end{eqnarray}

in \eqref{CasimirEnergy} together with the replacement of $\xi$ with $\xi_{l}= 2\pi\frac{k_{B}T}{\hbar}l$, the discrete Matsubara frequencies, takes into account the effect of temperature. We have defined $k_{B}$ to be Boltzmann's constant and the prime denotes that the term $l=0$ contains a prefactor $\frac{1}{2}$ compared to the other terms in the sum. This set of substitutions imply some assumptions which we now briefly discuss.\\
First of all we note that formally the dielectric function can also have temperature dependence through the parameters $\omega_{R_{i}}$,$\omega_{e_{i}}$ and $\gamma_{i}$  defined in \eqref{eq:phendiel}. Although to our knowledge there is no experimental data for topological insulators, these parameters are almost temperature independent for most dielectrics and so we exclude this effect in our analysis.
In addition, it is known that all dielectrics have non zero conductivity $\sigma(T)$ at $T \neq 0$ which modifies the dielectric function through $\epsilon(\omega, T) = \epsilon(\omega) + i4\pi\frac{\sigma(T)}{\omega}$. For dielectrics $\sigma(T)$ depends on the band gap $\Delta$: $\sigma(T)\sim \mathrm{exp}( -\frac{\Delta}{2k_{B}T})$.\\
Topological insulators such as Bi$_{2}$Se$_{3}$ have $\Delta \sim 0.3$eV \cite{XWQ09} and so we can neglect this contribution as a first approximation for temperatures below $T\sim 3\cdot10^{3}$ K which is enough for practical purposes. Nevertheless, it is possible to show \cite{BKMM09} that the inclusion of the finite conductivity term within the Lifshitz formalism leads to a violation of Nernst theorem which states that the entropy should tend to zero when $T\rightarrow 0$. Physically this is related to the appearance of a drift current, which leads to Joule heating and violates the condition of thermal equilibrium, necessary to apply the Lifshitz theory. Hence, the inclusion $\sigma(T)$ is therefore not justified within the Lifshitz theory and leads to large, unphysical thermal corrections which do not account for experimental data \cite{BKMM09}. Either way, we are forced to neglect this contribution. Other effects of temperature on the coating are discussed in section \ref{sec:conc}.\\
Having considered these points, we will assume that the formal substitution \eqref{eq:subs} is enough to take into account the effect of temperature on the Casimir effect in anisotropic topological insulators.

\subsection{Results at $T\neq0$ isotropic plates}

Once established the method to include temperature effects we analyze the case of two isotropic topological insulating plates described by (\ref{eq:phendiel}) at $T\neq0$. The Lifshitz equation (\ref{CasimirEnergy}) transforms for finite temperatures to:

\begin{equation}
\label{CasimirEnergytemp}
\frac{E_{c}(d)}{A} = k_{B}T{\sum^{\infty}_{l=0}}' \int \frac{d^2 {\bf k}_{\|}}{(2\pi)^2} \log \det \left[1 - {\bf R}_1 \cdot {\bf R}_2 e^{-2 k_3 d}\right].
\end{equation}

Where the transverse momentum $k_3$ is now evaluated at discrete frequencies $\xi_{l}$ such that $k_3=\sqrt{\bm{k}^{2}_{\|}+ \xi_{l}^2/c^2}$. It is convenient to define the adimensional temperature $\bar{T}$:

\begin{eqnarray}
\bar{T}\equiv \dfrac{2\pi k_{B}T}{\hbar\omega_{R}}.
\end{eqnarray}

For frequencies in the optical region, $\bar{T}=1$ correspond to temperatures $T\sim 2\cdot10^{3}$ K, while frequencies in the microwave region $\bar{T}=3.7$ correspond to temperatures $T\sim 300$K. Since we will be interested in the latter (the topological insulator TlBiSe$_{2}$ has a resonance frequency\cite{Chen10} at $\omega_{R}=56$cm$^{-1}$) our plots will range from $\bar{T}=0$ to $\bar{T}=10$. We note that the behavior is dominated by low frequency resonances as long as higher resonance frequencies have similar oscillator strengths in \eqref{eq:phendiel} (see auxiliary material in Ref.\cite{GC10}). Because of this, and the absence, to our knowledge, of broad frequency experimental data regarding the optical response of topological insulators (due mainly to their interest as thermoelectrics) we assume that this is the case and restrict to a one oscillator model throughout the paper. Nevertheless, as optical data becomes available, it is simple to implement these techniques and substitute the relevant parameters in the model since the overall assumptions will still hold (namely, the insulating behavior and the existence of a topological magnetoelectric polarizability).

We end here the discussion regarding the model and the theoretical tools and we now proceed to analyze both numerically and analytically the results obtained from integrating \eqref{CasimirEnergytemp}.

\subsubsection{Classical ($T\rightarrow\infty$) and quantum ($T\rightarrow 0$) limits}

To get some insight into the finite temperature behavior of the system one can ask what happens to the Casimir force in the (formal) limits when $T\rightarrow\infty$ and $T\rightarrow\ 0$. Trivially, if one first sets $T\rightarrow 0$ the expression (\ref{CasimirEnergy}) is recovered and the analysis of Ref. \cite{GC10} still holds. Hence a region of repulsion is expected to exist if the topological magnetoelectric polarizabilities of the plates have different signs.
For high temperature, $T\rightarrow\infty$ also referred to as the classical limit, all the terms in the $l$ sum above are exponentially suppressed except for the $l=0$ term \cite{BKMM09}. Thus, for isotropic plates, the dielectric function can be replaced by the zero frequency dielectric function. Let us derive the reflection coefficients in this case.\\
Directly from \eqref{eq:ReflectionMatricesTIcomplete}, and taking into account that $n^2_i=\epsilon_{i}$ the reflection coefficients at zero frequency take the form:

\begin{eqnarray}
r_s &=& \dfrac{1}{\Delta_{l=0}}(1-\epsilon(0)-\bar{\alpha}^2 +\sqrt{\epsilon(0)}\chi_{-}(0,\boldsymbol{k}_{\parallel})) \\
r_s &=& \dfrac{1}{\Delta_{l=0}}(-1+\epsilon(0)+\bar{\alpha}^2 +\sqrt{\epsilon(0)}\chi_{-}(0,\boldsymbol{k}_{\parallel})) \\
r_{sp} &=& \dfrac{2\bar{\alpha}}{\Delta_{l=0}} = r_{ps},
\end{eqnarray}

where $\Delta_{l=0} = 1+\epsilon(0)+\bar{\alpha}^2+\sqrt{\epsilon(0)}\chi_{+}(0,\boldsymbol{k}_{\parallel})$. From their definition \eqref{eq:Xipm}, the functions $\chi_{\pm}$ at zero frequency can be expressed in terms of $\epsilon(0)$ only ($\boldsymbol{k}_{\parallel}$ dependence cancels in this limit):

\begin{eqnarray}
\chi_{\pm} = \sqrt{\epsilon(0)}(1 \pm \dfrac{1}{\epsilon(0)})
\end{eqnarray}

Bringing together all simplifications the reflection coefficients (\ref{eq:ReflectionMatricesTIcomplete}) take the form:

%\begin{widetext}
\begin{eqnarray}
%\eqname{S.1}
\label{eq:ReflectionMatricesTIinftemp}
{\bf R}_i = \dfrac{1}{2(\epsilon(0)+1)+\bar{\alpha}^2}\left(
\begin{array}{cc}
  -\bar{\alpha}^2  &  2 \bar{\alpha} \\
  2 \bar{\alpha} & 2(\epsilon(0)-1)+\bar{\alpha}^2
\end{array} \right),
\end{eqnarray}
%\end{widetext}

\begin{figure}
\includegraphics[scale=0.9]{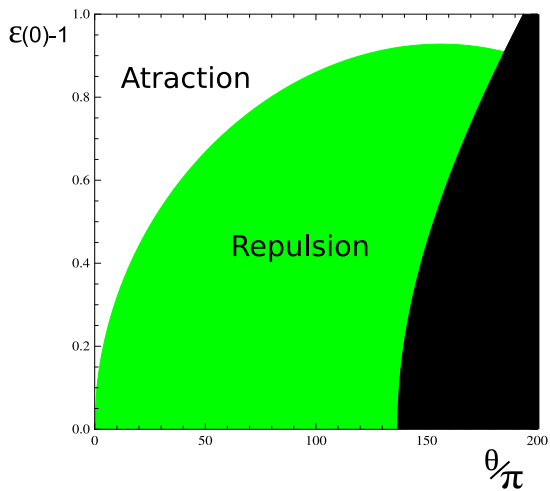}
\caption{\label{phasediagram}(color online) Attraction versus repulsion in the classical limit ($T\rightarrow\infty$) as a function of $\frac{\omega_{e}}{\omega_{R}}$ and $\frac{\theta}{\pi}$ where $\theta_{1}=-\theta_{2} \equiv\theta$. The white upper region shows attraction while the lower green region shows repulsion. The black region is a forbidden region for the parameters due to the positive energy condition $\frac{\alpha|\theta|}{\pi} < \sqrt{\epsilon(0)}$ (see text).}
\end{figure}

We emphasize that although similar in structure, these reflection coefficients are different to the ones obtained at zero temperature and  $d \rightarrow \infty$ in appendix \ref{App:Minimum} by rescaling the Casimir energy density with $d$.\\
One can introduce this expression in \eqref{CasimirEnergytemp} and study the behavior of the Casimir energy as a function of the two parameters of the model: $\epsilon(0)=1+\frac{\omega^2_{e}}{\omega^2_{R}}$ and $\theta_{1,2}$. In this case one can rescale the integration variables with $d$ and focus on the force defined as $F=-\partial_{d}E_{c}(d)$ which takes the form:

\begin{equation}\label{eq:classicalforce}
F_{cl}(d) = \dfrac{k_{B}T}{d^3\pi}f(\epsilon(0),\theta_{1},\theta_{2}),
\end{equation}

where $f$ is a complicated integral expression of the parameters (see appendix \ref{App:class} for details). It is evident from the functional form of \eqref{eq:classicalforce} with $d$ that in the classical limit the Casimir force is either attractive or repulsive for all distances, depending only on the sign of $f$. Thus, in contrast to the zero temperature case, no equilibrium point is expected.\\
When $\mathrm{sgn}(\theta_{1})=\mathrm{sgn}(\theta_{2})$ no repulsion is obtained for any values of the parameters, as expected from the results in Ref.\cite{GC10}. Whenever $\mathrm{sgn}(\theta_{1}) \neq \mathrm{sgn}(\theta_{2})$ it is possible to draw the diagram of Fig. \ref{phasediagram} where the attractive and repulsive behaviors are shown as a function of the two parameters for the particular case where $\theta_{1}=-\theta_{2} \equiv\theta$. To understand this diagram one first has to note that the topological magnetoelectric polarizability has an upper bound to satisfy the positive energy condition\cite{BHS68} for magnetoelectric materials. In the present context, this condition implies that $\frac{\alpha|\theta|}{\pi} < \sqrt{\epsilon(0)}$ which excludes the black region on the right of Fig. \ref{phasediagram}. The other two regions indicate attractive (white top region) or repulsive (green low region) behavior. From this diagram it is transparent that even in the classical limit of very high temperature there exists a region of parameters where complete repulsion is obtained. It is relevant to point out that a low dielectric response is needed for this to happen. High topological magnetoelectric polarizability also enhances the repulsive behavior.\\
Can one understand the origin of repulsion in the classical limit analytically? The answer is in fact affirmative. For small $\theta_{1,2}$ and $1-\epsilon(0)$, it is possible to write a closed analytical form for the Casimir force of the parameters although for the sake of clarity we leave this to appendix \ref{App:class}. In addition to this, we have included a discussion of the classical limit in terms of the Casimir energy density (see appendix \ref{App:class}) in the spirit of reference \cite{GC10} studying the relative strength of the diagonal and off-diagonal reflection coefficients. Both approaches lead to the same result: repulsion is possible in the classical limit for a certain range of parameters which are summarized in Fig.\ref{phasediagram}. Having studied these limiting cases we know proceed to the general case where $T\neq 0$.

\subsubsection{$T\neq 0$: the general case}

\begin{figure*}
\begin{minipage}{.32\linewidth}
\includegraphics[scale=0.39]{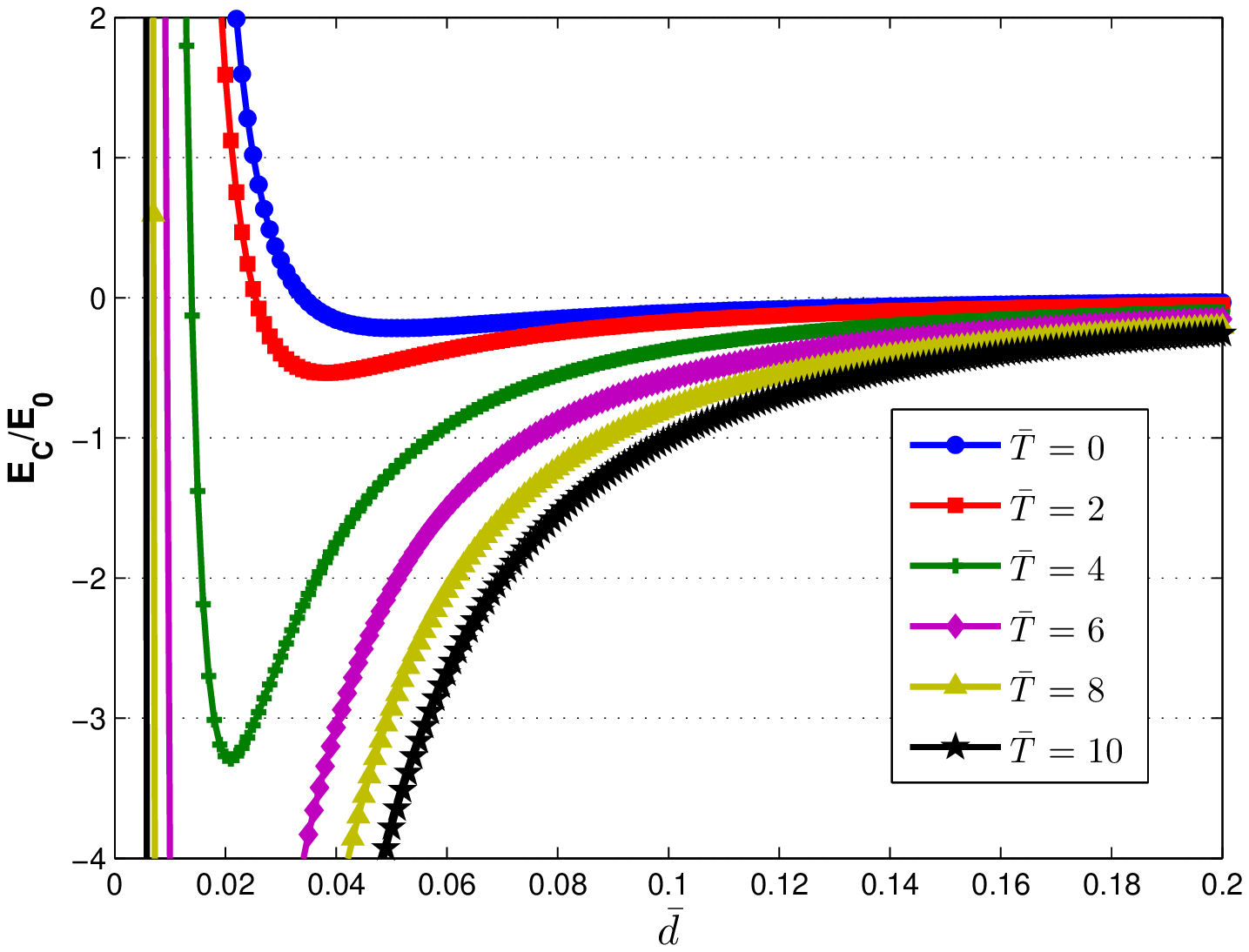}
(a)
\end{minipage}
\begin{minipage}{.32\linewidth}
\includegraphics[scale=0.40]{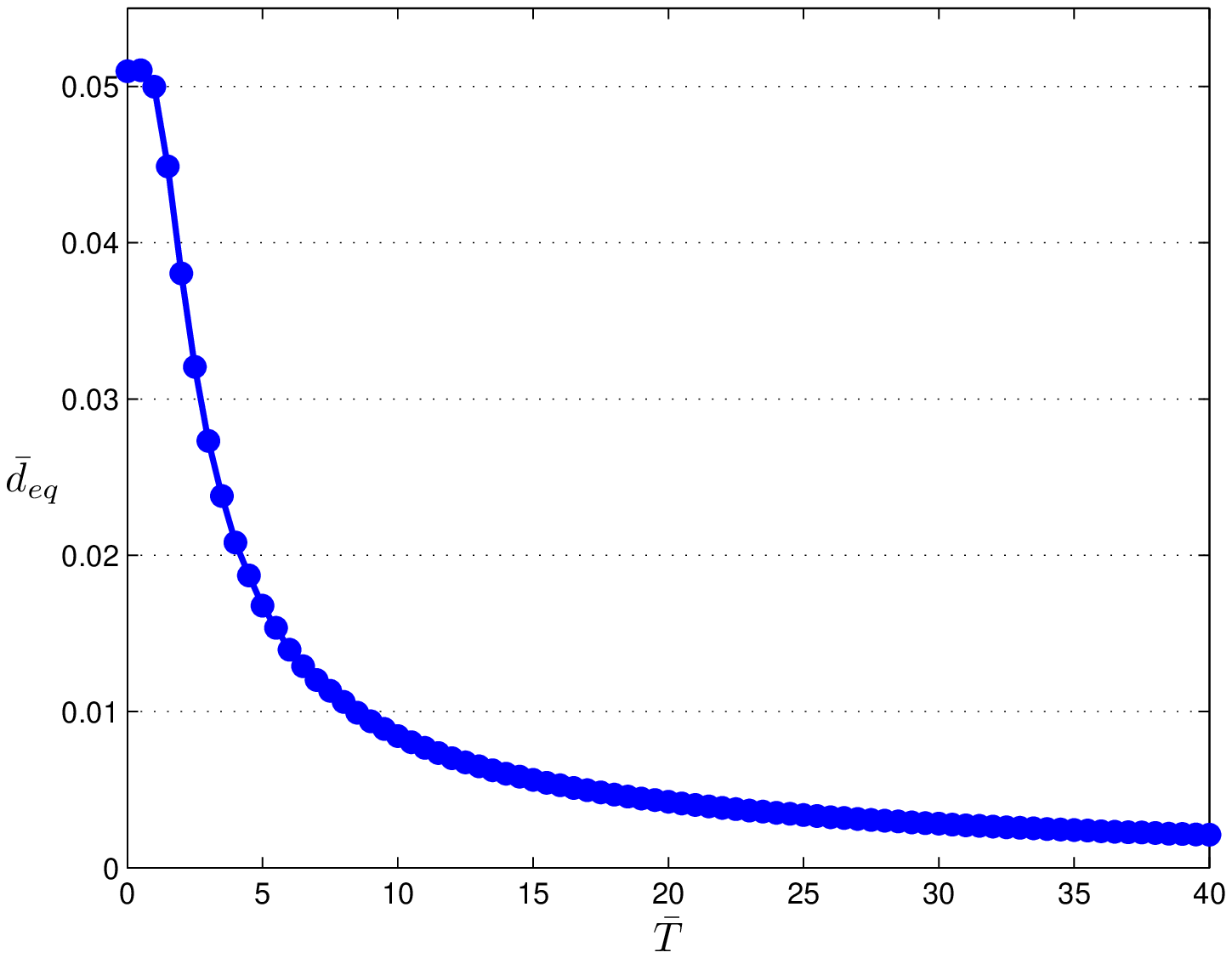}
(b)
\end{minipage}
\begin{minipage}{.32\linewidth}
\includegraphics[scale=0.40]{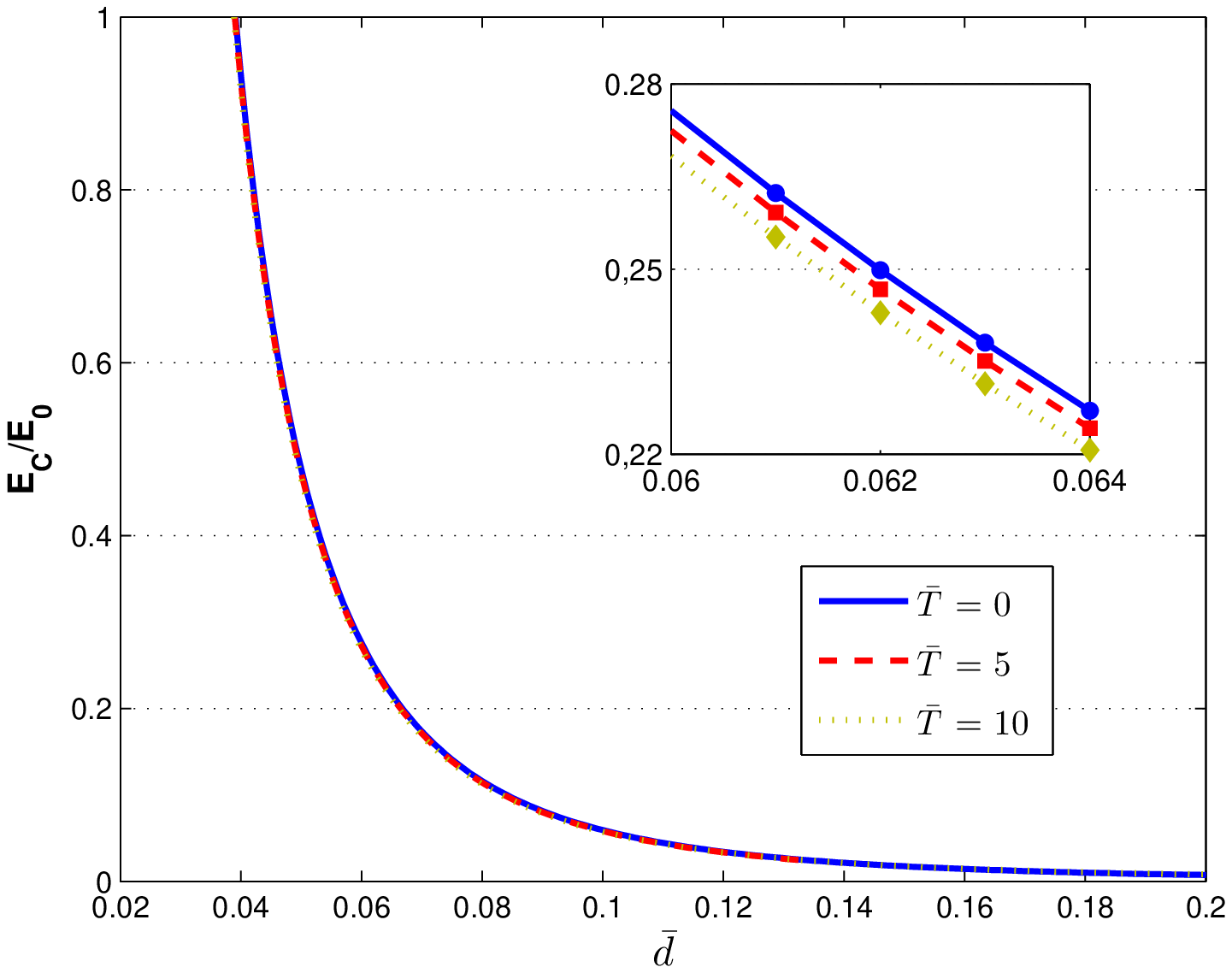}
(c)
\end{minipage}
\caption{\label{Fig:finiteT}(color online) (a) Casimir energy density (in units of $E_{0}= A\hbar c/(2\pi)^2(\omega_{R}/c)^3$) as a function of the dimensionless distance $\bar{d}$ for $\frac{\omega_{e}}{\omega_{R}}=0.45$ for temperatures ranging from $\bar{T}= 0-10$ and $\theta_{1}=-\theta_{2} = 3\pi$. Temperature pushes the repulsive behavior to shorter distances. (b) Equilibrium distance as a function of temperature. (c) Casimir energy density (in units of $E_{0}= A\hbar c/(2\pi)^2(\omega_{R}/c)^3$) as a function of $\bar{d}$ for $\frac{\omega_{e}}{\omega_{R}}=0.16$ for temperatures ranging from $\bar{T}= 0-10$. For values where there is repulsion, decreasing the temperature shifts the curve towards smaller distances as shown inside the inset figure.}
\end{figure*}

The effect of increasing temperature for a situation where $\mathrm{sgn}(\theta_{1})=-\mathrm{sgn}(\theta_{2})$ is presented in Fig.\ref{Fig:finiteT}. In general, temperature acts against repulsive behavior, driving it towards shorter distances as temperature increases. The evolution of the equilibrium position with respect to the adimensional temperature $\bar{T}$ is shown in Fig. \ref{Fig:finiteT}(b). For high values of $\bar{T}$ ($\bar{T}>5$) it is possible to complete equation \eqref{deqproptheta} (valid for small values of $\theta_{1,2}$) to write:

\begin{eqnarray}\label{deqpropthetaT}
\bar{d}_{eq}\propto  \dfrac{|\theta_{1}\theta_{2}|}{\bar{T}}.
\end{eqnarray}

This simple relation, extracted from numerical data, shows that for small topological magnetoelectric charge and high temperature both effects compete as regards to the equilibrium position.\\
One can summarize the results for this section as follows. In the classical limit, one can have either attraction or repulsion, regardless of the distance. Starting with attraction at the classical limit, i.e $\bar{T} \rightarrow \infty$, decreasing temperature will develop a deeper minimum which will shift to longer distances as shown in Fig.\ref{Fig:finiteT}(a) and (b). Consistent with the discussion presented the end of section \ref{sec:T=0}.B, in some special cases where the magnetoelectric is bigger than a critical value one can end with repulsion at $T=0$ without a minimum.\\
Starting from repulsion in the classical limit will not develop a minimum when decreasing $\bar{T}$ but it will shift the curve of repulsive behavior towards larger distances to end with repulsion in the quantum limit. This is shown in Fig.\ref{Fig:finiteT}(c). It must be noted that for this to happen, a very low $\epsilon(0)$ is needed when $\theta\sim\pi$ and as $\theta$ is increased, higher $\epsilon(0)$ are capable of giving rise to the complete repulsive behavior. Therefore we expect this not to be the relevant situation for real topological insulators, which have low $\theta$, but rather the one presented in Fig. \ref{Fig:finiteT}(a). An extension of this analysis could be relevant for other magnetoelectric materials such as Cr$_{2}$O$_{3}$ that can present higher axion coupling as well as more general magnetoelectric couplings \cite{Obu08}.\\
We note for completeness that when both topological magnetoelectric couplings have equal signs, temperature acts to increase attraction as expected from ordinary dielectrics.\\
We finish this section with a technical comment regarding the commutativity of the limits. We have discussed the high and low temperature limits, and studied the distance dependence numerically. At this point, one might ask whether the limits commute, i.e. if one evaluates first the  $d\rightarrow 0$ or $d\rightarrow \infty$ limit first and then consider $T\rightarrow 0$ and $T\rightarrow \infty$, would the result remain as stated above?
The fact is that in some cases, the limits commute while in others they do not. In particular, is clear from expression \eqref{CasimirEnergytemp} that the limits $d\rightarrow \infty$ and $T\rightarrow \infty$ will commute. It is easy to show that, in either order, both result into the same reflection coefficients given by \eqref{eq:ReflectionMatricesTIinftemp}. For the case when $d\rightarrow 0$ and $T\rightarrow 0$ a similar situation occurs and the limits also commute. The interesting case is however when $d\rightarrow 0$ and $T\rightarrow \infty$ (or viceversa). If one expands the exponential function, then taking the temperature to infinity breaks down the expansion and this limit is not reconciled with the $T\rightarrow \infty$, $d\rightarrow 0$ which typically gives attraction. Although it might seem surprising at first, one should recall other physical situations where a similar situation happens, one particularly impressive example taking place when calculating the DC or minimal conductivity of graphene where the order of evaluating the limits is crucial (for a recent review of the problem see Ref.\cite{SAHR10} section IV.A). Nevertheless, it should be noted that, with the numerical results in mind, there is a way around the puzzle by imposing that we take the limits in such a way that the product $Td$ is a constant. In this way, the $l=0$ is the leading term in the sum \eqref{CasimirEnergytemp} and we recover the reflection coefficients \eqref{eq:ReflectionMatricesTIinftemp} relevant for high temperature, making the limits commute.

\section{\label{sec:uniaxial}Casimir force between topological insulators with uniaxial anisotropy at $T=0$}

In this section we again set the temperature to zero and focus on the effect that uniaxial anisotropy has on the repulsive behavior.

\subsection{Anisotropy in topological insulators}

For a given material its linear response to an electromagnetic wave, or in turn, its reflection coefficients defined in (\ref{CasimirEnergy}) depend strongly on the dielectric tensor $\epsilon_{ij}$, the magnetic susceptibility tensor $\mu_{ij}$ and the magnetoelectric tensor $\theta_{ij}$. Since we are dealing with non magnetic topological insulators we will assume that $\mu_{ij}=\delta_{ij}$, although we shall derive the reflection coefficients for the general case in which the susceptibility is considered.

Crystal symmetry defines the other two tensors. Due to time reversal symmetry, for topological insulators the magnetoelectric tensor can only have the form $\theta_{ij}=\theta\delta_{ij}$ with $\theta=\pi$  (mod($2\pi$)) for topological insulators and $\theta =0$ for ordinary insulators \cite{EJV10}. It is interesting to note that for general magnetoelectric crystals, other magnetoelectric couplings can be present in the action \eqref{eq:lagrangean}, and $\theta_{ij}$ could have a more elaborate form \cite{LL8, EJV10} which could in principle affect the sign of the Casimir force. We will restrict ourselves to the case of a constant isotropic magnetoelectric tensor $\theta_{ij}=\theta\delta_{ij}$, relevant for crystalline topological insulators \cite{EJV10}.\\
The form of $\epsilon_{ij}$ is also determined by crystal symmetry and one can distinguish the cubic, uniaxial and biaxial crystals\cite{LL8}. The simplest case, the cubic structure, has $\epsilon_{ij}=\epsilon(\omega)\delta_{ij}$, which is just the isotropic case considered in previous sections. Although interesting as a first approximation \cite{GC10}, this model is incomplete for topological insulators. The prototypical examples of topological insulating crystals Bi$_2$Se$_3$, Bi$_2$Te$_3$ , Sb$_2$Te$_3$ \cite{Chen10,C09,X09} or the novel Thallium based III-V-VI$_{2}$ compounds \cite{Zhang10,Chen10,LMW10a,LMW10b} have rhombohedral symmetry, which fall into the category of uniaxial crystals and hence we will focus on this case, leaving biaxial crystals aside. Uniaxial crystals have a definite optical axis which coincides with the principal crystal axis. The dielectric tensor can thus be written in the form: $\epsilon_{ij}=\mathrm{diag}(\epsilon_{\perp},\epsilon_{\perp},\epsilon_{z})$, where the optical axis is aligned with the $z$ axis. In all our calculations, the optical axis of the two Casimir plates are supposed to be aligned so that no torque is induced in the system \cite{BKMM09}.

\subsection{Fresnel coefficients for topological insulators with uniaxial anisotropy at $T=0$}

Fresnel coefficients are derived from Maxwell's equations by imposing continuity relations for the normal components of $\bf{D}$ and tangential components of $\bf{H}$ as discussed above for the isotropic case (see appendix \ref{App:coef}). After some tedious but straightforward work one obtains the reflection matrices for an anisotropic topological insulator-vacuum interface:

\begin{widetext}
\begin{eqnarray}
\label{eq:anisotropicandTI}
\mathbf{R}_{i} =\dfrac{1}{\Delta_{an}}\left(
\begin{array}{cc}
\left(\mu_{\perp}k_{z}-q'\right) \left(\epsilon_{\perp}k_{z}+q'' \right)-q''k_{z}\mu_{\perp}\bar{\alpha}^2  & 2\bar{\alpha}q''\mu_{\perp}k_{z}\\
  2\bar{\alpha}q''\mu_{\perp}k_{z} & \left( \epsilon_{\perp}k_{z}-q''\right) \left(\mu_{\perp}k_{z}+q' \right)+q''k_{z}\mu_{\perp}\bar{\alpha}^2
\end{array} \right),
\end{eqnarray}
\end{widetext}

where $\Delta_{an}= \left( \mu_{\perp}k_{z}+q'\right) \left(\epsilon_{\perp}k_{z}+q'' \right)+q''k_{z}\mu_{\perp}\bar{\alpha}^2 $, $k^2_{z} = \dfrac{\omega^2}{c^2}-k^{2}_{\parallel}$, $q'^{2} = \dfrac{\omega^2}{c^2}\epsilon_{\perp}\mu_{\perp}-k^2_{\parallel}\dfrac{\mu_{\perp}}{\mu_{z}}$, $q''^{2} = \dfrac{\omega^2}{c^2}\epsilon_{\perp}\mu_{\perp}-k^2_{\parallel}\dfrac{\epsilon_{\perp}}{\epsilon_{z}}$ and $\bar{\alpha} = \alpha\theta/\pi$.
To compute the Casimir energy, one should define these in the imaginary frequency axis and perform the substitution $\omega = i\xi$.
We stress that these reduce to the ordinary anisotropic coefficients presented in appendix \ref{App:coef} when $\bar{\alpha}=0$. They also reduce to the isotropic reflection coefficients \eqref{eq:ReflectionMatricesTIcomplete} when $\mu$ and $\epsilon$ are isotropic.\\
To investigate the effect of anisotropy it is necessary to model the dielectric response of the material. In general, insulators can be described by its resonance frequencies, hence we model $\epsilon_{\perp}$ and $\epsilon_{z}$ with a sum of oscillators:

\begin{eqnarray}\label{eq:phendiel1}
\epsilon_{z}(i\xi) &=& 1+\sum_{i}\dfrac{\omega_{e_{z},i}^2}{\xi^2+\omega^2_{R_{z,i}}+\gamma_{z,i}\xi}, \\
\label{eq:phendiel2}
\epsilon_{\perp}(i\xi) &=& 1+\sum_{i}\dfrac{\omega_{e_{\perp},i}^2}{\xi^2+\omega^2_{R_{\perp,i}}+\gamma_{\perp,i}\xi}.
\end{eqnarray}

We have evaluated the dielectric function at imaginary frequencies, as demanded by equation \eqref{CasimirEnergy}. The parameters $\omega_{e_{\perp,i}}$ and $\omega_{e_{z,i}}$ indicate the strength of the oscillator while $\omega_{R_{\perp,i}}$ and $\omega_{R_{z,i}}$ account for the resonance frequencies of the oscillators. Finally, we include the damping parameters $\gamma_{\perp,i}$ and $\gamma_{z,i}$ for completeness although they do not play a mayor role on Casimir physics.

A multiple oscillator model does not significantly alter the general behavior of the Casimir force\cite{GC10} and so, in order to clarify the effects that different parameters have on the Casimir energy density we will restrict the models \eqref{eq:phendiel1} and \eqref{eq:phendiel2} to one oscillator models. In this scheme we will investigate two aspects: (i) the effect of modifying the relative strength of the parallel component against the perpendicular i.e. $\omega_{e_{z}}$ over $\omega_{e_{\perp}}$ and (ii) the effect of the relative movement of the position of the resonance frequencies $\omega_{R_{\perp}}$ and $\omega_{R_{z}}$.

\subsection{Effect of the relative strength of the parallel component against the perpendicular component}

To study this effect we simplify equations \eqref{eq:phendiel1} and \eqref{eq:phendiel2} to unmask the effect under study (i.e. point (i) above). Redefining the parameters in units of $\omega_{R_{\perp}}$:

\begin{eqnarray}\label{eq:phendiel1p1}
\epsilon_{z}(i\xi) &=& 1+\dfrac{\omega_{e_{z}}^2}{\xi^2+ 1 +\gamma\xi}, \\
\label{eq:phendiel2p1}
\epsilon_{\perp}(i\xi) &=& 1+\dfrac{\omega_{e_{\perp}}^2}{\xi^2+ 1 +\gamma\xi}.
\end{eqnarray}

\begin{figure*}
\begin{minipage}{.32\linewidth}
\includegraphics[scale=0.39]{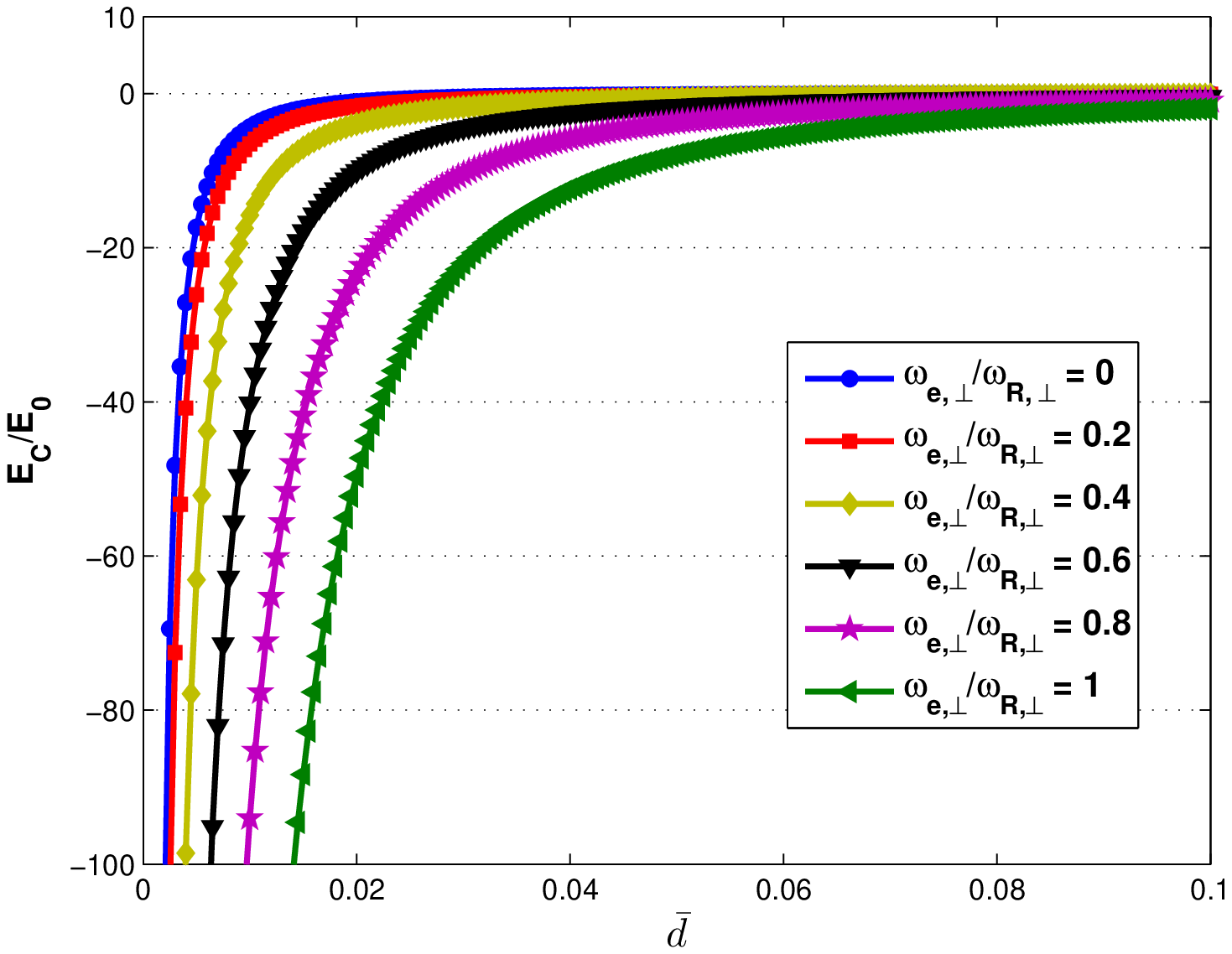}
(a)
\end{minipage}
\begin{minipage}{.32\linewidth}
\includegraphics[scale=0.4]{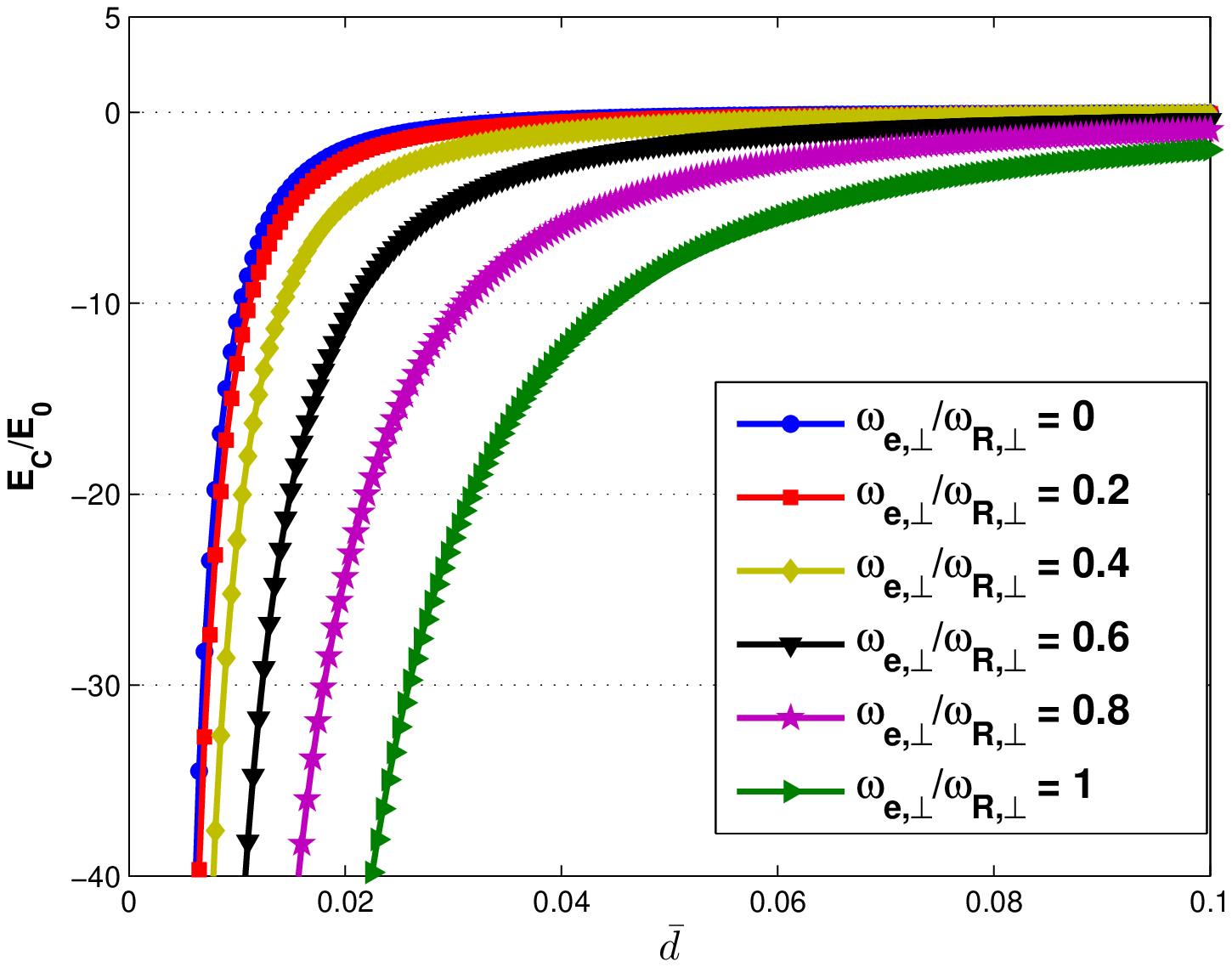}
(b)
\end{minipage}
\begin{minipage}{.32\linewidth}
\includegraphics[scale=0.4]{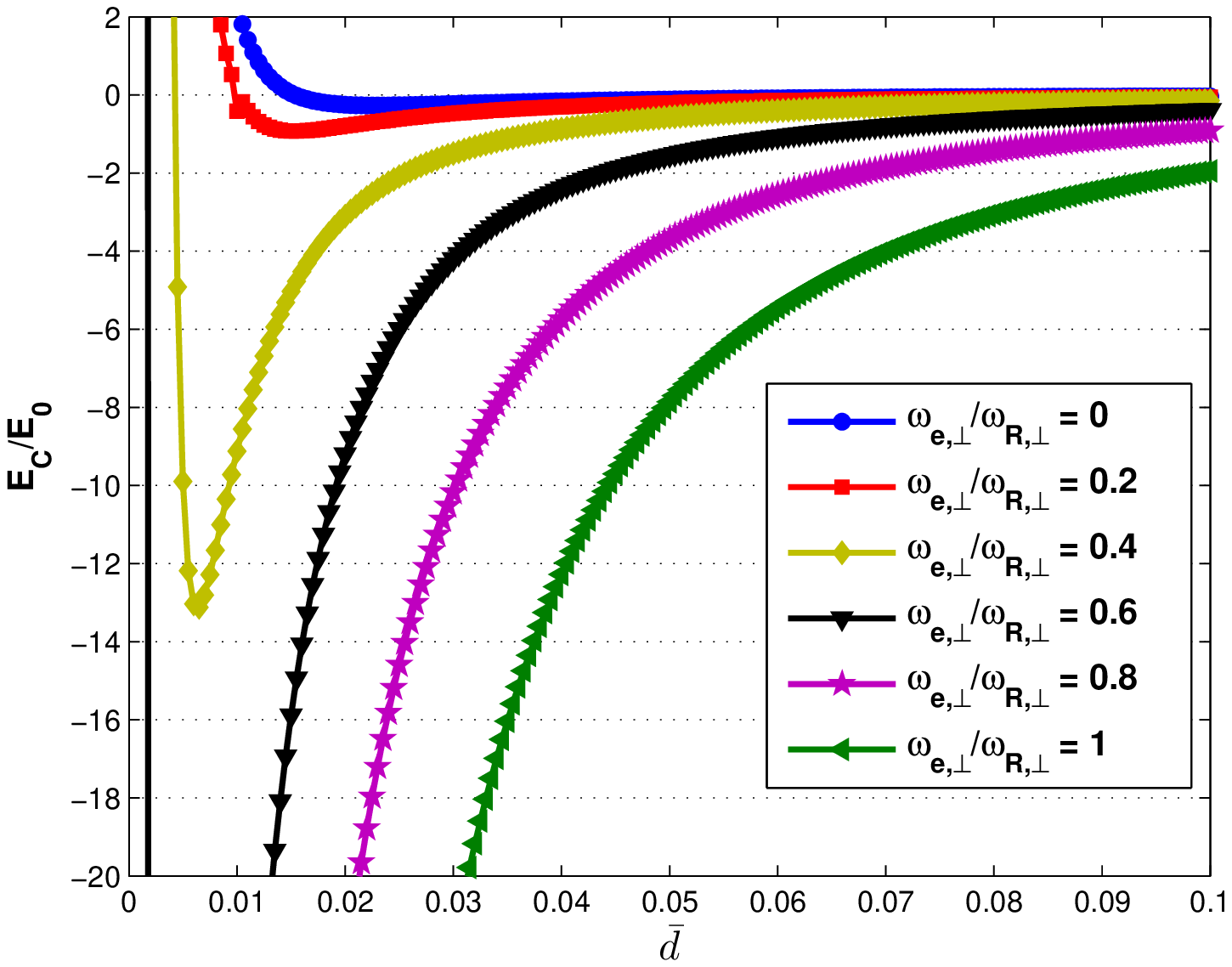}
(c)
\end{minipage}
\caption{\label{Fig1} (color online)Effect of changing the relative oscillator strengths between parallel and perpendicular components of the dielectric function on the Casimir energy density (in units of $E_{0}= A\hbar c/(2\pi)^2(\omega_{R_{z}}/c)^3$) as a function of the dimensionless distance $\bar{d}$ with (a) $\theta_{1,2}=0$ and (b) $\theta_{1}=\theta_{2}=\pi$ (c) $\theta_{1}=-\theta_{2}=\pi$. (a) Attraction is obtained as no magnetoelectric term is included. (b) Both magnetoelectric couplings have the same sign and so no repulsion occurs. (c) Repulsion appears at short distances. Increasing $\omega_{e_{\perp}}$ makes the minimum shift to lower distances while increasing its depth. Therefore, repulsion is favored by the condition $\omega_{e_{z}}>\omega_{e_{\perp}}$ }
\end{figure*}

In this case we have chosen $\omega_{R_{\perp}}=\omega_{R_{z}}$ and $\gamma\equiv\gamma_{R_{\perp}}=\gamma_{R_{z}}=0.01$ in units of $\omega_{R_{\perp}}$ . To study the effect of changing the relative strength of the dielectric components we may fix $\omega_{e_{z}}=0.45$ (in units of $\omega_{R_{\perp}}$) and change $\omega_{e_{\perp}}$ in the interval $0-1$. The results are shown in Fig.\ref{Fig1}. We have fixed $\theta_{1}=-\theta_{2}=0$ (i.e. the topologically trivial anisotropic case) in Fig. \ref{Fig1}(a), $\theta_{1}=\theta_{2}=\pi$ in Figure \ref{Fig1}(b) and $\theta_{1}=-\theta_{2}=\pi$ in Fig. \ref{Fig1}(c).

Fig. \ref{Fig1}(a) is a reference figure where only anisotropy is studied with $\theta=0$. In this case, attraction occurs since $\theta_{i}=0$ and there is no mixing of polarizations since the diagonal component in \eqref{eq:anisotropicandTI} is zero (see appendix \ref{App:coef}). As $\omega_{e_{\perp}}$ grows the tendency is to favor attraction and to increase the absolute value of the Casimir energy at a given distance. In Fig. \ref{Fig1}(b) the magnetoelectric terms have equal sign and so no repulsion is obtained just as it is expected from the isotropic case\cite{GC10}. When $\theta_{1}$ and $\theta_{2}$ have opposite sing, $\theta_{1}=-\theta_{2}=\pi$, the Casimir energy develops a minimum, shown in Fig. \ref{Fig1}(c). The actual distance at which the minimum appears is shifted to shorter distances as $\omega_{e_{\perp}}$ is increased. This enables to draw the conclusion that for the minimum to shift to larger distances, or in other words, for the repulsive behavior to be enhanced, the material in question has to favor the strength of oscillators in the direction parallel to the optical axis. \\
Physically, repulsion in this system is due to the mixing of polarizations\cite{GC10}. Thus, suppression of repulsion can be traced back to the relative suppression of the off-diagonal terms in \eqref{eq:anisotropicandTI} versus the diagonal terms just as in the anisotropic case studied in \cite{GC10}. When increasing $\epsilon_{\perp}$ fixing $\epsilon_{z}$ to a constant, the off-diagonal terms vanish, confirming the numerical analysis.

\subsection{Effect of the relative position of the oscillator frequencies}

To study this case, the dielectric functions of both components can be written as:

\begin{eqnarray}\label{eq:phendiel1p2}
\epsilon_{z}(i\xi) &=& 1+\dfrac{\omega_{e}^2}{\xi^2+ 1 +\gamma\xi}, \\
\label{eq:phendiel2p2}
\epsilon_{\perp}(i\xi) &=& 1+\dfrac{\omega_{e}^2}{\xi^2+ \beta +\gamma\xi}.
\end{eqnarray}

\begin{figure*}
\begin{minipage}{.32\linewidth}
\includegraphics[scale=0.39]{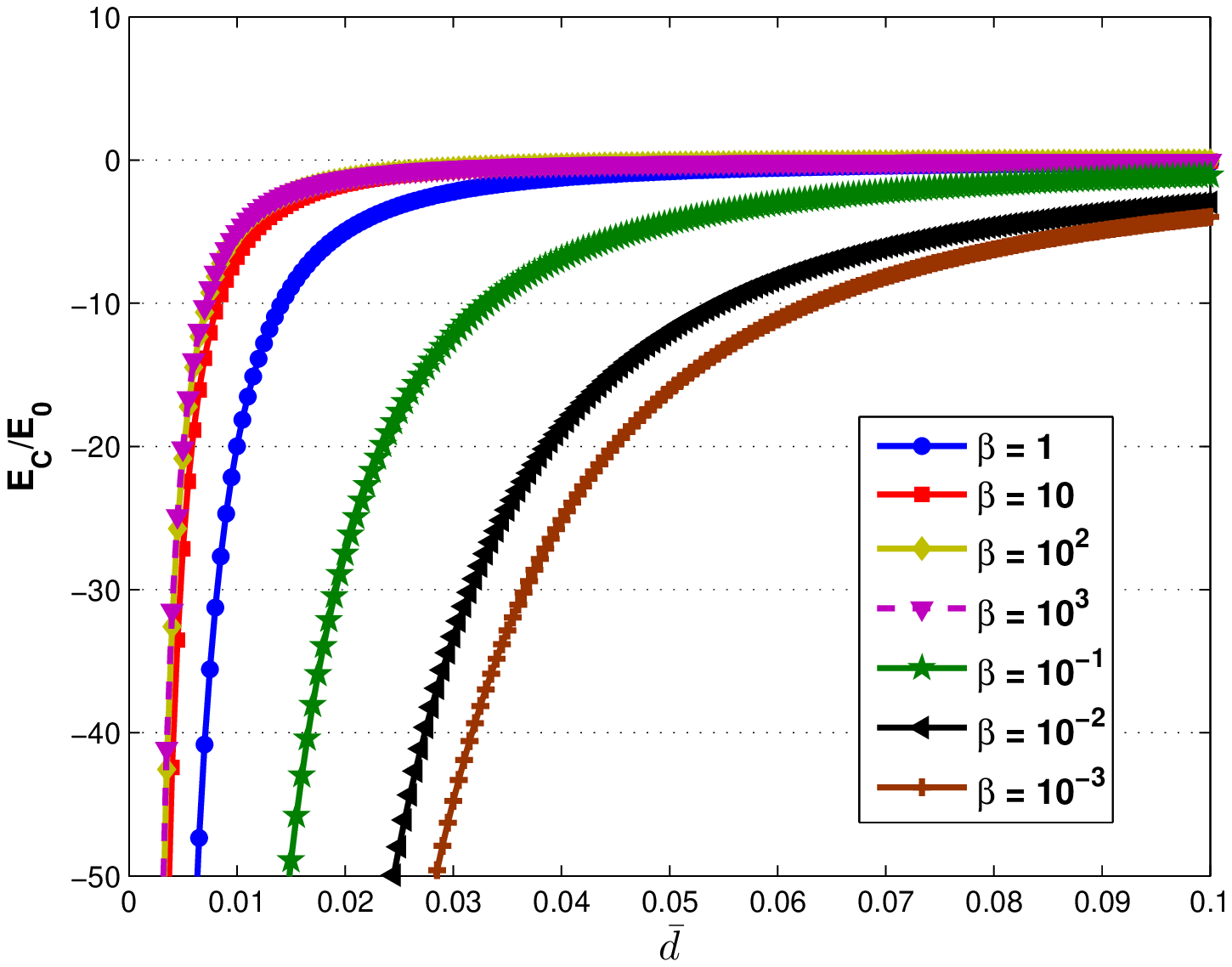}
(a)
\end{minipage}
\begin{minipage}{.32\linewidth}
\includegraphics[scale=0.4]{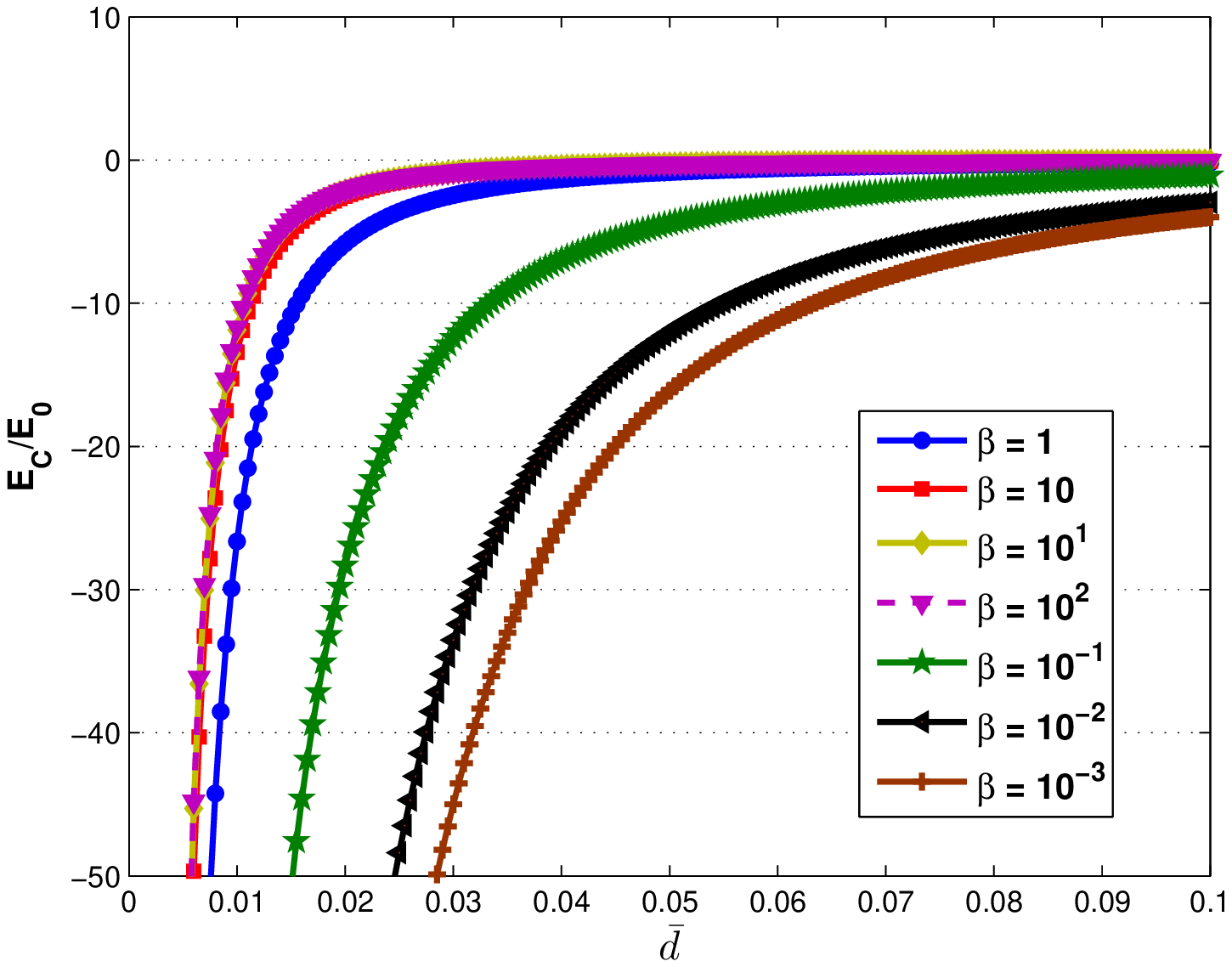}
(b)
\end{minipage}
\begin{minipage}{.32\linewidth}
\includegraphics[scale=0.4]{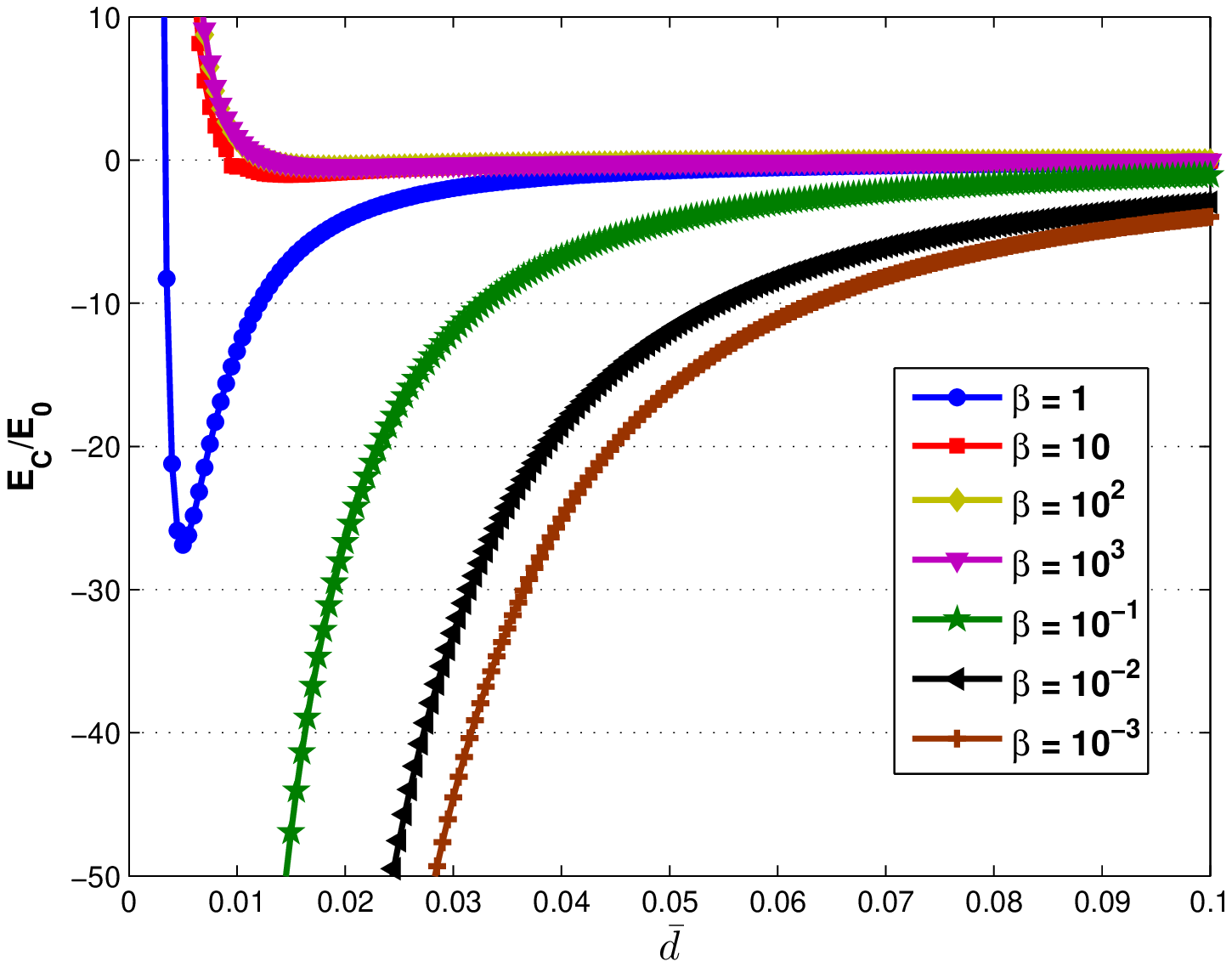}
(c)
\end{minipage}
\caption{\label{Fig2}(color online) Effect of changing the relative positions of the oscillator resonances between parallel and perpendicular components of the dielectric function on the Casimir energy density (in units of $E_{0}= A\hbar c/(2\pi)^2(\omega_{R_{z}}/c)^3$) as a function of the dimensionless distance $\bar{d}$ with (a) $\theta_{1,2}=0$ and (b) $\theta_{1}=\theta_{2}=\pi$ (c) $\theta_{1}=-\theta_{2}=\pi$. (a) Attraction is obtained as no magnetoelectric term is included. (b) Both magnetoelectric couplings have the same sign and so no repulsion occurs. (c) Increasing $\omega_{R_{\perp}}$ makes the minimum shift to higher distances while decreasing its depth. Therefore, repulsion is favored by increasing $\beta$.(see text)}
\end{figure*}

In this case we have chosen $\omega_{e_{\perp}}=\omega_{e_{z}}\equiv \omega_{e}=0.45$ and as before  $\gamma=0.01$ (in units of $\omega_{R_{\perp}}$). The parameter $\beta\equiv \frac{\omega^2_{R_{\perp}}}{\omega^2_{R_{z}}}$ determines the relative position between both oscillators: for $\beta>1$, $\epsilon_{\perp}$ has a higher resonance frequency than $\epsilon_{z}$ and for $\beta<1$ the opposite situation occurs. Again we have fixed $\theta_{1}=-\theta_{2}=0$ in Fig. \ref{Fig2}(a), $\theta_{1}=\theta_{2}\equiv\theta_{b}=\pi$ in Fig. \ref{Fig2}(b) and $\theta_{1}=-\theta_{2}\equiv\theta_{c}=\pi$ in Fig. \ref{Fig2}(c).

As in the previous case, Fig. \ref{Fig2}(a) a reference figure where only parameter $\beta$ is modified. We have chosen the values of $\beta$ ranging from $\beta=10^{-3}$ to $\beta=10^{3}$, so a broad range of situations are represented. In this situation, attraction occurs since $\theta_{i}=0$ and there is no mixing of polarizations. As $\beta$ grows the tendency is to favor attraction and to increase the absolute value of the Casimir energy at a given distance. In Fig. \ref{Fig2}(b) the magnetoelectric terms have equal sign and so no repulsion is obtained. Fig. \ref{Fig2}(c) shows how the minimum is shifted to lower distances as $\omega_{R_{\perp}}$ is increased, i.e. $\beta$ is increased. Hence, for the repulsive behavior to be favored, the material in question has to favor low resonance frequencies in the parallel direction, while high frequencies in the perpendicular direction. This is consistent with the behavior discussed above where suppression of $\epsilon_{\perp}$ enhances repulsion. In this case, increasing $\beta$ suppresses $\epsilon_{\perp}$ and favors repulsion \footnote{This effect is analogous to the effect discussed in Ref. \cite{GC10} (auxiliary information) where a higher oscillator frequency was studied and seen to enhance repulsion.\\}. \\
We summarize the results of this section in what follows. From the numerical results one can readily infer what conditions are necessary for repulsion to be observed, given that the topological magnetoelectric terms of the two materials have opposing signs. For the repulsive behavior to appear at the largest distances possible, one should search for a material where the direction parallel to the optical axis has a bigger oscillator strength than the corresponding strength of the oscillator in the parallel direction $\omega_{e_{z}}>\omega_{e_{\perp}}$. Further enhancement of the repulsive behavior can be achieved if the oscillator resonances in the direction perpendicular to the optical axis are at higher frequency than resonances for the parallel direction, i.e $\omega_{R_{z}}>\omega_{R_{\perp}}$. In both situations, $\epsilon_{\perp}$ is suppressed relative to $\epsilon_{z}$ and the mixing of polarizations is maximized.\\
With these numerical calculations we have established the main directions to enhance repulsive behavior with topological insulators at low temperature. We now discuss the effect of temperature on these findings.

\section{\label{sec:uniaxialandtemp}Casimir force between topological insulators with uniaxial anisotropy at $T\neq0$}

\begin{figure*}
\begin{minipage}{.49\linewidth}
\includegraphics[scale=0.5]{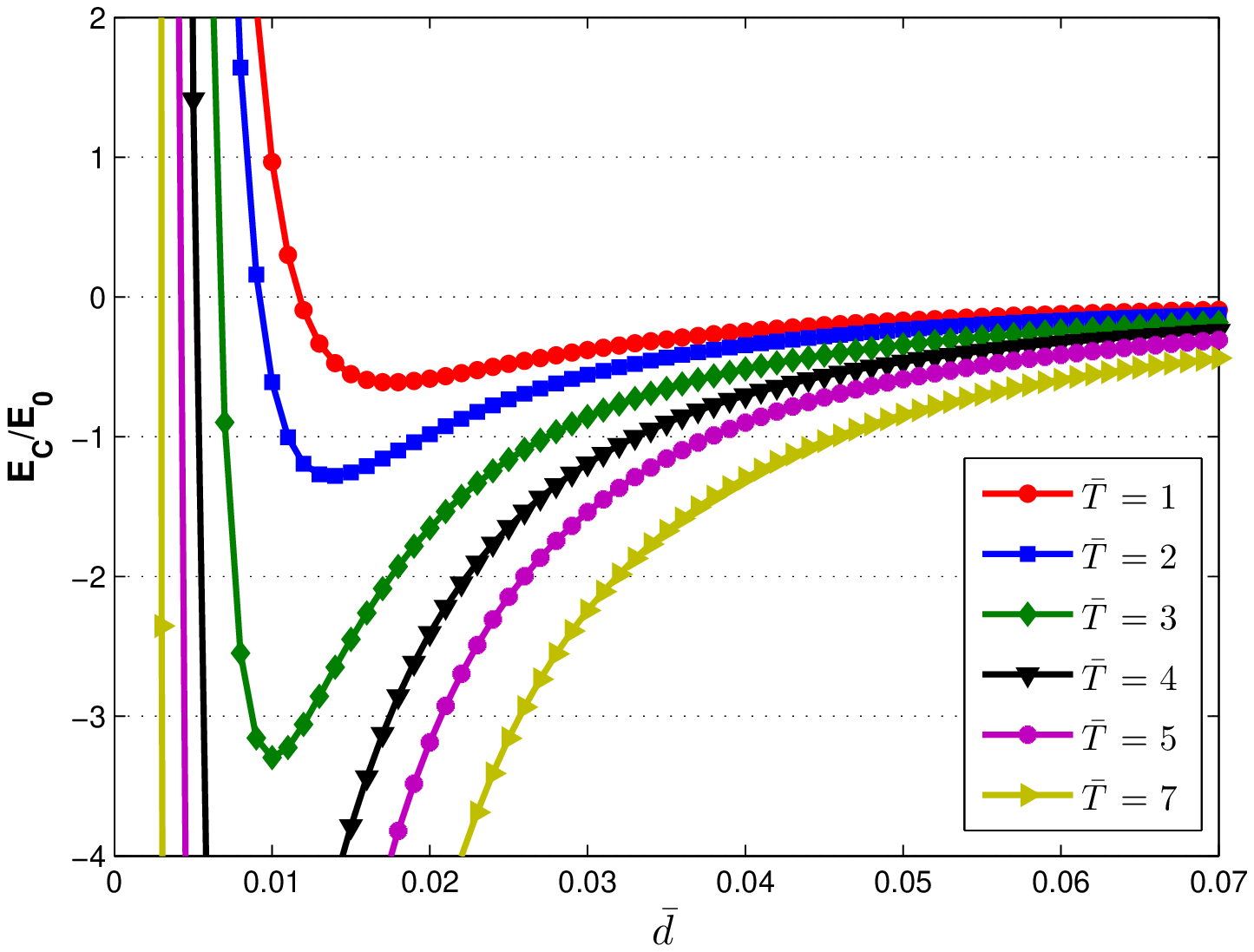}
(a)
\end{minipage}
\begin{minipage}{.49\linewidth}
\includegraphics[scale=0.5]{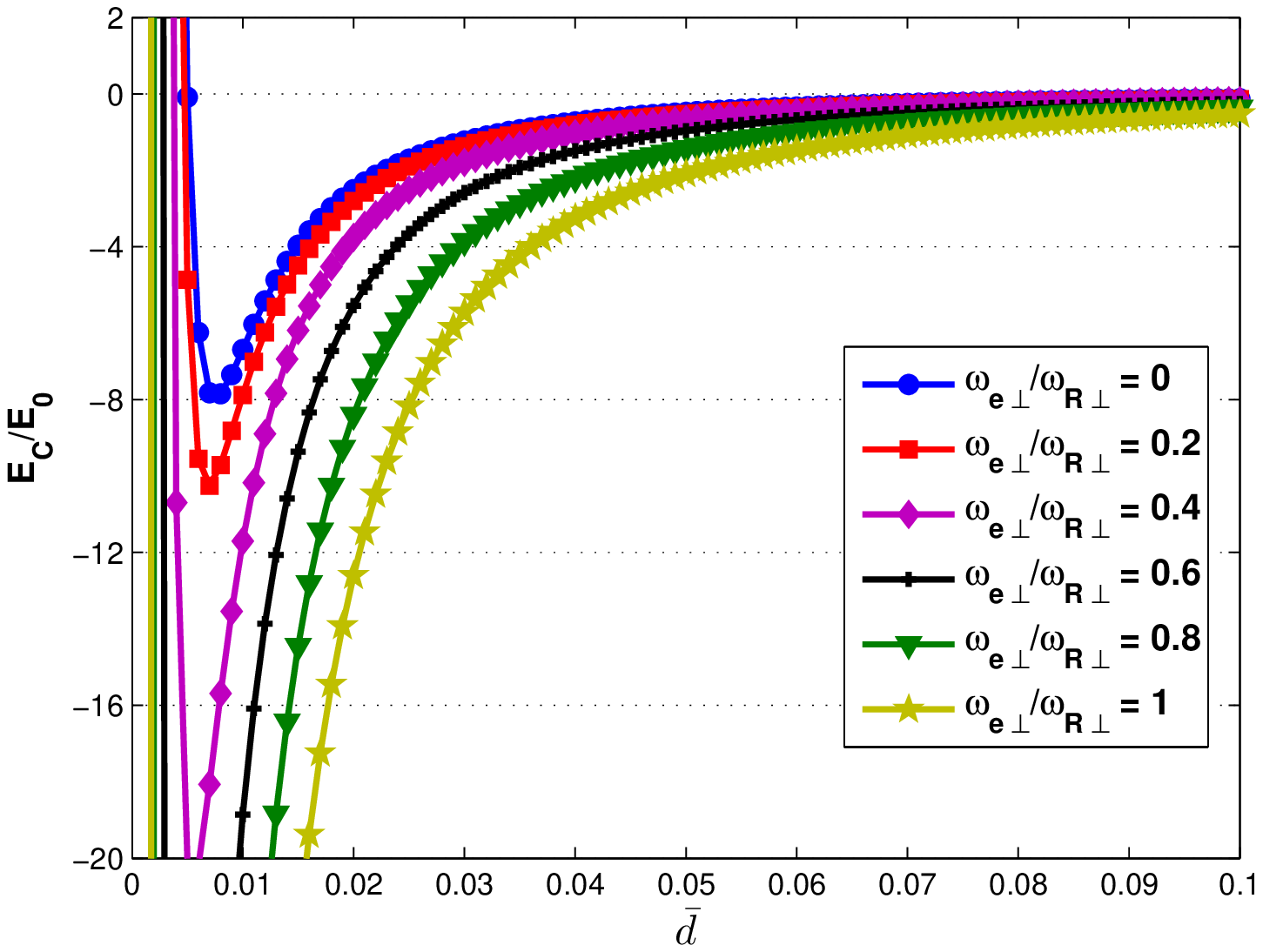}
(b)
\end{minipage}
\caption{\label{Fig:Anistemp}(color online) Casimir energy density (in units of $E_{0}= A\hbar c/(2\pi)^2(\omega_{R,z}/c)^3$) and $\theta_{1}=-\theta_{2}=\pi$ including anisotropy and temperature effects: (a) Effect of changing the temperature from $\bar{T}=0 - 7$ with $\omega_{e,z}=0.45$, $\omega_{e,\perp}=0.3$ and $\beta=5$. (b) Effect of varying $\omega_{e,\perp}=0-1$ with $\bar{T}=5$, $\omega_{e,z}=0.45$ and $\beta=5$.}
\end{figure*}

The result of including the two effects subject of this work, temperature and anisotropy together is shown in Fig \ref{Fig:Anistemp}. To illustrate that the repulsive behavior is still present we study two different cases. In the first of them, corresponding to Fig. \ref{Fig:Anistemp}(a) we vary the adimensional temperature from $\bar{T}=0$ to $\bar{T}=7$ while fixing the uniaxial anisotropy parameters and $\theta_{1}=-\theta_{2}=\pi$. In particular we fix the oscillators strengths to $\omega_{e_{z}}=0.45$, $\omega_{e_{\perp}}=0.3$ and the relative position of the resonances to $\beta=5$, i.e. the perpendicular oscillator is shifted by a factor of five with respect to $\omega_{R_{z}}$. As expected by previous sections, the figure proves that temperature works against the repulsive behavior, shifting repulsion to shorter distances as temperature is increased. In a second situation we illustrate the effect of varying anisotropy at a given temperature. In this case, we fix the temperature to $\bar{T}=5$, the oscillator strength in the parallel direction to $\omega_{e_z}=0.45$ and $\beta=5$ while varying $\omega_{e_{\perp}}$ from $\omega_{e_{\perp}}=0$ to $\omega_{e_\perp}=1$. The results show that enhancing perpendicular response, even at finite temperature, favors repulsive behavior. Consistently with previous sections, equal signs of $\theta_{1,2}$ the system returns to attraction at all distances.

\section{\label{sec:conc}Discussion and Conclusions}

In this work we have studied the effect of anisotropy and temperature on the repulsive Casimir effect with three dimensional topological insulators. We have found that the repulsive behavior is still present even when including these effects. In particular, increasing temperature tends to reduce repulsion, similar to what is expected from different situations such as metamaterials \cite{Rosa08}. On the other hand, enhancing the optical response in the direction parallel to the optical axis of the topological insulator works in favor of repulsion. This can be achieved either by increasing the relative strength of the oscillator in the parallel direction or by searching for resonance frequencies which suppress the perpendicular response. We have also determined that both effects together still allow for repulsion. All these effects can be understood as an enhancement of the off-diagonal terms in the reflection coefficients, which favors polarization mixing and thus repulsion, whenever the signs of the topological magnetoelectric terms are opposite in both plates.  \\
Although these specific conclusions are reached, it is difficult to draw conclusions for real materials since the optical response of topological insulators still needs further experimental characterization. In particular, values for the optical parameters are still lacking, in contrast with other materials which are traditionally useful for measuring the Casimir force \cite{BKMM09,Palik98}. Nevertheless, in this work, we have aimed also to establish the theoretical framework for handling anisotropy in topological insulators in the context of their optical properties, which to our knowledge was absent in the literature. The reflection coefficients for anisotropic materials can also be helpful in achieving the optical characterization of these materials and could be relevant for metrology purposes, as suggested in previous works\cite{TM10,MQDZ10}. This theoretical work on anisotropy might also be used as a starting point for the study of the Casimir force in situations involving other magnetoelectric materials.\\
A final issue left out in this work is the effect of temperature on the magnetic coating. Even though magnetic interactions where shown to be irrelevant \cite{GC10}, temperature will have an effect on the magnetization. In particular, the magnetic moments will fluctuate with increasing temperature and destroy ferromagnetism when the system is over the Curie temperature ($T_{c}$) of the coating. This temperature depends on the specific coating material which should be an insulating or semiconducting ferromagnet with as high as possible Curie temperature.
Although ferromagnets are usually metallic, examples of insulating ferromagnets exist, for example oxides such as the rare earth oxide $Eu0$ with Curie temperature close to $70$K \cite{H78}, and have been synthesized and used in the past, such as the ferromagnetic insulator GdN. The latter has been shown to have a Curie temperature close to $60$K \cite{H78}, it can be grown as a thin film of thickness $\sim6$nm\cite{XC96} and could be a suitable candidate for the ferromagnetic covering. \\
In conclusion we expect that the tools and behaviors studied in this work might lead towards a complete optical characterization of topological insulators, a first step towards Casimir like experiments in which the description of factors such as temperature and anisotropy should be relevant.

\section*{Acknowledgements}

Two of us, A.G.G. and A.C., are indebted to M.A.H. Vozmediano for fruitful conversations and encouraging discussions. A.G.G. has enjoyed conversations with T. Emig and acknowledges financial support from MICINN through grant No. FIS2008-00124 (FPI Program). P.R.-L. acknowledges helpful discussions with R. Brito. P.R.-L. research supported by projects MODELICO, MOSAICO and a FPU MEC grant. A.C. acknowledges Alianza 4 Universidades Program for funding.

\appendix

\section{\label{App:Minimum}Proof of the existence of the minimum at $T=0$: Large and short distance limits of the Casimir Energy density:}

In this appendix we give detailed information and expressions of the some results presented in \cite{GC10} so this work is self-contained. We prove the existence of the minimum by discussing the high and low frequency limits of the Casimir energy density. The Casimir energy stored between the plates is given by:

\begin{equation}
\label{CasimirEnergys}
\frac{E_{c}(d)}{A\hbar} = \int_0^{\infty} \hspace{-1pt} \frac{d\xi}{2\pi} \int \frac{d^2 {\bf k}_{\|}}{(2\pi)^2} \log \det \left[1 - {\bf R}_1 \cdot {\bf R}_2 e^{-2 k_3 d}\right] ,
\end{equation}

where ${\bf R}_i$ is the $2$x$2$ matrix defined in \eqref{eq:ReflectionMatricesTIcomplete}.
To prove the existence of the minimum we assume that the reflection coefficients are given by the expression \eqref{eq:ReflectionMatricesTIcomplete} and that the dielectric function satisfies the following two analytical properties: 1) finite dielectric permittivity at zero frequency ($\epsilon(0)<\infty$) and 2) high frequency transparency: $\epsilon(\omega)\rightarrow 1 $ when $\omega\rightarrow \infty$. For an insulator one can assume the dielectric function is given in \eqref{eq:phendiel}. For simplicity let's assume that:

\begin{equation}\label{eq:phendiels}
\epsilon(i\xi)=1+\dfrac{\omega_{e}^2}{\xi^2+\omega^2_{R}+\gamma_{R}\xi} ,
\end{equation}

although the derivation does not depend on the analytical form of the dielectric function as long as it fulfills the mentioned conditions (the proof holds for more than one oscillator). For analytical traceability we assume the particular situation where $\theta_{1}=-\theta_{2}$ (where labels $1$ and $2$ identify the Casimir plates).\\
The first step is to rescale $E_{c}(d)$ with $d$. If in the expression for $E_{c}(d)$, $\xi$ and ${\bf k}_{\|}$ are rescaled to contain $d$, tan overall factor $1/d^3$ appears in front of the integral. The reflection matrices are to be evaluated now at the rescaled frequency and momenta $\xi /d$ and ${\bf k}_{\|}/d$. Thus, all of the $d$ dependence can be transferred to the reflection matrices and the overall $1/d^3$ prefactor. The expression for the energy is:

\begin{widetext}
\begin{equation}
\label{CasimirEnergyres}
\frac{E_{c}(d)}{E_{0}} = \dfrac{1}{d^3}\int_0^{\infty} \hspace{-1pt} d\xi \int d^2 {\bf k}_{\parallel} \log \det \left[1 - {\mathbf{R}_1\left[\dfrac{\xi}{d},\dfrac{\mathbf{k}_{\parallel}}{d}\right]} \cdot {\mathbf{R}_2\left[\dfrac{\xi}{d},\dfrac{\mathbf{k}_{\parallel}}{d}\right]} e^{-2 k_3}\right] ,
\end{equation}
\end{widetext}

where $k^{2(r)}_{3} = \xi^2+k_{\parallel}^{2}$ is defined through the rescaled variables (which include $c$ the speed of light in vacuum and $\omega_R$) and $E_{0}= A\hbar c/(2\pi)^2(w_{R}/c)^3$ where $A$ is the area of the plates. The key point is that reflection matrices are evaluated at these rescaled variables, with a rescaled dielectric function:

\begin{equation}\label{eq:phendielres}
\epsilon(i\xi/d)=1+\dfrac{\left( \dfrac{\omega_{e}}{\omega_{R}}\right) ^2}{\left( \dfrac{\xi}{d}\right) ^2+ 1+\dfrac{\gamma_{R}}{\omega_{R}}\dfrac{\xi}{d}}.
\end{equation}

The two conditions imposed to $\epsilon(i\xi)$ ensure that the reflection matrices are not singular in the limits $\xi/d \rightarrow 0$ and $\xi/d \rightarrow \infty$. Hence, $E_{c}(d\rightarrow 0) \rightarrow \pm\infty$ and $E_{c}(d\rightarrow \infty) \rightarrow 0$. \\
The way the integrand approaches these limits determines the sign of $E_{c}(d)$. For instance, if the integrand is positive at small distances and negative at large distances, necessarily a minimum exists at an intermediate distance. When $\theta_{1}=-\theta_{2}$, this is exactly what happens (unless $\epsilon(0)=1$, where both limits are positive and hence long range repulsion is obtained), as it will be shown in what follows.\\
We now evaluate the integrand in \eqref{CasimirEnergys}. When $\theta_{1}=-\theta_{2}= \theta$ the reflection matrices describing both topological insulators can be written as:

\begin{eqnarray}
\label{ReflectionMatricesTIpm}
{\bf R}_{\pm} = \left[
\begin{array}{cc}
   r_{s} (i \xi, {\bf k}_{\|}) &  \pm r_{sp} (i \xi, {\bf k}_{\|}) \\
  \pm r_{sp} (i\xi, {\bf k}_{\|}) &  r_{p} (i \xi, {\bf k}_{\|})
\end{array} \right] .
\end{eqnarray}

Introducing this inside \eqref{CasimirEnergys} the integrand follows:

\begin{widetext}
\begin{eqnarray}\nonumber
I = \log \det \left[1 - \mathbf{R}_{+}\cdot\mathbf{R}_{-} e^{-2 k^{(r)}_3}\right] = \log\left[ 1 + e^{-2 k^{(r)}_3} \left(2 r_{sp}^2-r_{p}^2-r_{s}^2\right) + e^{-4 k^{(r)}_3} \left(r_{sp}^2-r_{p}r_{s}\right)^2\right],
\end{eqnarray}
\end{widetext}

Notice that the last term although always positive, will play no role in what follows since it is always suppressed over the first term (note that $r_{s},r_{p},r_{sp} < 1$).

\subsection*{The limit of short distances:}

For $d \rightarrow 0$ and using the high frequency transparency of the dielectric function we now show that $|r_{i}|<<|r_{sp}|$ ($i=s,p$):
Notice first that the denominator $\Delta$ defined in \eqref{eq:ReflectionMatricesTIcomplete} is common to all terms so it cannot play a role on the relative magnitude of the coefficients. We hence study the behavior of $\chi_{-}$ at small distances, given by:

\begin{eqnarray}
\label{eq:Ximsmall}
 \chi_{-}\left(\dfrac{\xi}{d},\dfrac{\mathbf{k}_{\parallel}}{d}\right) =
 \dfrac{\xi^2 + \mathbf{k}_{\|}^2 - \left(\xi^2 + \frac{\mathbf{k}_{\|}^2}{n_{2}^2}\right)}{\sqrt{\left( \xi^2 + \mathbf{k}_{\|}^2\right) \left(\xi^2 + \frac{\mathbf{k}_{\|}^2}{n_{2}^2}\right)}}.
\end{eqnarray}

Remembering that at small distances (large frequencies) we have transparency, $n_{2}^{2}(\xi/d)= \epsilon(\xi/d) \rightarrow 1$ then we see that $\chi_{-}\rightarrow 0 $ and:

\begin{equation}\nonumber
r_{s} = \dfrac{-\bar{\alpha}^2}{2+\bar{\alpha}^2 + \chi_{+}} = -r_{p},
\end{equation}

and

\begin{equation}\nonumber
r_{sp} = \dfrac{2\bar{\alpha}}{2+\bar{\alpha}^2 + \chi_{+}},
\end{equation}

since the first are of order $\mathcal{O}(\alpha^2)$ and the second are of order $\mathcal{O}(\alpha)$ (remember that $\bar{\alpha}$ is proportional to the fine structure constant $\alpha$) we have that $|r_{i}|<<|r_{sp}|$ ($i=s,p$).  Hence the integrand is positive (since the integrand has the form $I=\ln(1+A)$ where $A > 0$) and so $E_{c}(d\rightarrow 0) \rightarrow +\infty$.

\subsection*{The limit of large distances:}

When $d \rightarrow \infty$ the reflection coefficients take the form

\begin{equation}\nonumber
r_{s} = \dfrac{1-\epsilon (0)-\bar{\alpha}^2 + \sqrt{\epsilon (0)}\chi_{-}}{1+\epsilon(0)+\bar{\alpha}^2 + \sqrt{\epsilon(0)}\chi_{+}},
\end{equation}

for the diagonal part (with a similar expression for $r_{p}$) and

\begin{equation}\nonumber
r_{sp} =  \dfrac{2\bar{\alpha}}{1+\epsilon(0)+\bar{\alpha}^2 + \sqrt{\epsilon(0)}\chi_{+}},
\end{equation}

for the off diagonal. We have defined the quantity  $\epsilon(0) \equiv 1+ \left( \frac{w_{e}}{w_{R}}\right)^2$.
In this case, depending on the values of $\epsilon(0)$ different behaviors emerge. Since $\epsilon(0) \geq 1$ we can distinguish to extreme limits, one where $\epsilon(0)=1$ and the other with $\epsilon(0)>>1$.\\
When $\epsilon(0)=1$ one can see that we return to the previous case since the quantity $\chi_{-}$ in this limit also goes to zero. Hence $|r_{sp}|>>|r_{i}|$ ($i=s,p$) is satisfied for all distances and $E_{c}(d)$ is always positive .Therefore, using the fact that that $E_{c}(d) \rightarrow 0$ when $d \rightarrow \infty$  and that $E_{c}(d) \rightarrow +\infty$ when $d \rightarrow 0$ we deduce that there is no minimum and that the force is always repulsive. This was confirmed by the numerical calculations presented in \cite{GC10}. \\
When $ \epsilon(0) >> 1$, we see that the opposite condition, $|r_{s,p}|>>|r_{sp}|$, is satisfied. Even in the worst case when $\chi_{-}$ is smallest, $\epsilon(\xi)$ in $r_{s}$ is always larger than $2\alpha$ in $r_{sp}$. The integrand at large distances is a negative quantity (since the integrand now has the form $I=\ln(1-B)$ with $B > 0$) and so $E_{c}(d)$ approaches to zero from negative values ($E_{c}(d) \rightarrow -\infty$ for $d \rightarrow \infty$).
Since at small distances $E_{c}(d)\rightarrow +\infty$ we conclude that there must be a minimum at an intermediate distance $ 0 < d_{m} < \infty$, since the function must cross the $d$ axis.\\
Notice that when $\epsilon(0)$ is strictly infinity we recover the case of an ideal metal where $r_{s,p} = \mp 1$ and $r_{sp} = 0$, with attraction at all distances.\\
To sum up, as we increase $\epsilon(0)$ from one, a minimum develops at an intermediate distance $d_{m}$. This distance shifts to lower values as we increase $\epsilon(0)$ until, at $\epsilon = \infty$ we recover the ideal metal case where complete attraction occurs.\\
In the case where $\theta_{1}=\theta_{2}$ the signs inside $I$ change to give:
\begin{widetext}
\begin{eqnarray}\nonumber
I = \log\left[ 1 - e^{-2 k^{(r)}_3} \left(2 r_{sp}^2+r_{p}^2+r_{s}^2\right) + e^{-4 k^{(r)}_3} \left(r_{sp}^2-r_{p}r_{s}\right)^2\right].
\end{eqnarray}
\end{widetext}
The predominant term inside the logarithm is always negative and hence the integrand is always negative. Therefore at small distances $E_{c}(d)\rightarrow -\infty$ and at large distances $E_{c}(d)\rightarrow 0$ approaching this limit from negative values, recovering attraction at all intermediate distances.\\
Finally, it can be checked by analogous methods, that the case where one $\theta$ is zero and the other one is not (dielectric - topological insulator case) results in Casimir attraction for all distances.

\section{\label{App:class}Classical limit: analytical expressions}

The Casimir force at non zero temperature is defined as $F=-\partial_{d}E_{c}(d)$ with $E_{c}(d)$ given by \eqref{CasimirEnergytemp}:
\begin{widetext}
\begin{equation}\label{Finite_T_adimensional_force_TI}
\dfrac{F\left(T,d\right)}{A} = - \frac{\hbar\omega_{R}^{4}}{4\pi^{2}c^{3}}\bar{T}{\sum_{l=0}^{\infty}}'\int_{0}^{\infty}dk_{\parallel}k_{\parallel}k_{3}\text{Tr}\left[\left(1 - {\bf R}_1 \cdot {\bf R}_2 e^{-2 k_3 \bar{d}}\right)^{-1}{\bf R}_1 \cdot {\bf R}_2 e^{-2 k_3 \bar{d}}\right],
\end{equation}
\end{widetext}

where $k_3=\sqrt{k_{\parallel}^{2} + \bar{T}^{2}l^{2}}$ and $R_{i}$ are the reflection coefficients of each plate. In this expression the momentum variable is rescaled in units of $\omega_{R}$ and the integral depends only on $\bar{d}$ and $\bar{T}$. As discussed in the main text, at high temperature, only the $l=0$ term contributes since the rest are exponentially suppressed. Keeping only this term in the integral one can recast this expression as:

\begin{equation}
F_{cl}(d) = \dfrac{k_{B}T}{d^3\pi}f(\epsilon(0),\theta_{1},\theta_{2}).
\end{equation}

Integrating this expression for different values of $\theta_{1,2}$ and $\epsilon(0)$ one obtains the diagram in Fig.\ref{phasediagram} (whenever $\theta_{1,2}$ have opposite signs but equal magnitude $\theta$).\\
In order to derive analytical results further approximations must be considered. In the rest of this appendix we will assume that we are always in the region of small magnetoelectric coupling $\bar{\alpha}<<1$, which is justified as long this coefficient is of the order of the fine structure constant, assumption to be expected in real topological insulators. Hence, for small topological magnetoelectric couplings we can evaluate the integral to give:
\begin{widetext}
\begin{equation}
F_{cl}(d) = -\dfrac{k_{B}T}{8d^3\pi}\left[Li_{3}\left(\frac{(\epsilon(0) - 1)^{2}}{(\epsilon(0) + 1)^{2}}\right) + \left(\bar{\alpha}_{1}^{2} + \bar{\alpha}_{2}^{2} + 2\bar{\alpha}_{1}\bar{\alpha}_{2}\frac{\epsilon(0) + 1}{\epsilon(0) - 1}\right)\frac{1}{(\epsilon(0) - 1)(\epsilon(0) + 1)}Li_{2}\left(\frac{(\epsilon(0) - 1)^{2}}{(\epsilon(0) + 1)^{2}}\right)\right],
\end{equation}
\end{widetext}

where $\bar{\alpha}=\frac{\theta\alpha}{\pi}$ and $\mathrm{Li}_{n}$ is the polylogarithm of order $n$. When $\epsilon(0)\rightarrow 1$, the force takes the simple form:

\begin{equation}\label{eq:appforcehighTlimit}
F_{cl}(d)=-\dfrac{k_{B}T}{16d^3\pi}\bar{\alpha}_{1}\bar{\alpha}_{2}.
\end{equation}

From this last expression it is transparent that one can still get repulsion in the classical limit when the signs of $\theta_{1}$ and $\theta_{2}$ are opposite.\\
As anticipated in the main text one can analyze the Casimir energy density in the classical limit in the spirit of Ref. \cite{GC10}, that is, exploring the relative magnitudes of the diagonal and off-diagonal parts of the reflection coefficients.
In this case, we assume for simplicity that both topological magnetoelectric polarizabilities have the same magnitude but can have equal or opposite signs, i.e. $\mathrm{sgn}(\theta_{1})=\pm\mathrm{sgn}(\theta_{2})$. If we introduce these reflexion matrices in expression (\ref{CasimirEnergytemp}) and take the high temperature limit ($l=0$), after some algebra one arrives to:

\begin{equation}
\label{CasimirEnergyhightemp}
\frac{E^{\pm}_{c}(d)}{A} = \dfrac{k_{B}T}{4\pi d^2} \int^{\infty}_{0} k_{\|}d k_{\|} \log \left[1-A_{\pm}e^{-2k_{\|}}+Be^{-4k_{\|}}\right],
\end{equation}

where the $\pm$ corresponds to each of the two cases $\mathrm{sgn}(\theta_{1})=\pm\mathrm{sgn}(\theta_{2})$ and we have rescaled the integral variables with the distance. The functions $A_{\pm}$ and $B$ are defined as:

\begin{eqnarray}\nonumber
A_{\pm} &=& \frac{1}{D}(4(\epsilon(0)-1)^2+(4(\epsilon(0)-1)\pm 8)\bar{\alpha}^{2}+2\bar{\alpha}^{4}),\\
\nonumber
B &=& \dfrac{\bar{\alpha}^4}{D},
\end{eqnarray}

where $D\equiv \left(2(\epsilon(0)+1)+\bar{\alpha}^2\right)^2$. Note that these two coefficients depend on the zero frequency dielectric function $\epsilon(0)$ and the absolute value of the topological magnetoelectric polarizability $|\theta|$ as well as its sign in the case of $A_{\pm}$. These functions govern the sign of the integral. Function $B$ plays a secondary role since it is exponentially suppressed. It is clear, that when $\epsilon(0)\rightarrow 1$ the integrand takes the form $I=\mathrm{ln}(1-x)$ where $x$ is positive or negative depending on $A_{\pm}$. For $A_{-}$, i.e. $\mathrm{sgn}(\theta_{1})=-\mathrm{sgn}(\theta_{2})$ and $\epsilon(0)\to 1$, $x<0$ so $I>0$ which makes $E\propto\frac{k_{B}T}{4\pi d^2}$ and the force is repulsive since a negative (positive) force, or equivalently a positive (negative) slope of $E_{c}(d)$, corresponds to attraction (repulsion) of the plates. \\
Indeed, when one neglects the second term inside the logarithm in \eqref{CasimirEnergyhightemp} the integral can be expressed in terms of a Polylogratihm:

\begin{equation}\label{eq:appenergyhighTlimit}
\frac{E^{\pm}_{c}(d)}{A} = - \frac{k_{B}T}{16\pi d^{2}}\mathrm{Li}_{3}(A_{\pm})
\end{equation}

The function $\mathrm{Li}_{3}(A_{\pm})$ has the sign of $A_{\pm}$. Therefore, when $A_{\pm}>0$($<0$) we have attraction (repulsion). Using the fact that $\mathrm{Li}_{3}(x)\sim x +\mathcal{O}(x^2)$ when $x\to 0$ and that $A_{-} \sim -\frac{\bar{\alpha}^2}{2}$ when $\epsilon(0)\to 1$ and to leading order in $\bar{\alpha}^2$ we can readily obtain the Casimir force from \eqref{eq:appenergyhighTlimit} which gives:

\begin{equation}
F_{cl}(d)=\dfrac{k_{B}T}{16d^3\pi}\bar{\alpha}^{2}.
\end{equation}

This equation is exactly equation \eqref{eq:appforcehighTlimit} when $\bar{\alpha}_{1}=-\bar{\alpha}_{2}=\bar{\alpha}$. Consistently, both approaches are equivalent and confirm our numerical results, where we found that even at high temperature repulsion can occur. Nevertheless, for more realistic situations where higher values of $\epsilon(0)$ are expected, there is a competition between parameters, which is represented in Fig.\ref{phasediagram}. Only high values of $\theta$ can compete with large enough $\epsilon(0)$ and thus we expect attraction to occur at high temperatures.

\section{\label{App:coef}Fresnel coefficients for topological insulators with uniaxial anisotropy}

In this appendix, the Fresnel coefficients for topological insulators with uniaxial anisotropy will be discussed. First, we will briefly review the derivation of the reflection coefficients for an ordinary dielectric - vacuum interface so that the extension to the topological insulator - vacuum interface is more transparent.

\subsection*{Fresnel coefficients for uniaxial material plates}

As an illustrative case, before adding the axionic term, we proceed to solve Maxwell's equations for an uniaxial material which will demonstrate the procedure to follow in the next section. We start with an uniaxial material with its optical axis parallel to the plate's normal. The corresponding dielectric permittivity tensor is given by $\epsilon_{ij}=\mathrm{diag}(\epsilon_{\perp},\epsilon_{\perp},\epsilon_{z})$ and, for completeness, the magnetic susceptibility tensor will be included as $\mu_{ij}=\mathrm{diag}(\mu_{\perp},\mu_{\perp},\mu_{z})$. The subindex $z$ indicates that the optical axis is aligned with the $z$ axis, chosen also to be the surface normal.
The procedure is to solve Maxwell's equation in vacuum, then inside the uniaxial media, and finally impose boundary conditions to determine the reflection amplitudes.
The first step is therefore to solve Maxwell's equations in vacuum by proposing a plane wave solution of the form (see for instance Ref.\cite{RDM08})

\begin{eqnarray}\nonumber
\mathbf{E}_{in} &=& \left(A_{\perp}\mathbf{y}+A_{\parallel}\dfrac{c}{\omega}(k_{z}\mathbf{x}-k_{x}\mathbf{z}) \right) e^ {i(k_{x}x+k_{z}z-\omega t)},\\
\nonumber
\mathbf{H}_{in} &=& \left(A_{\parallel}\mathbf{y}-A_{\perp}\dfrac{c}{\omega}(k_{z}\mathbf{x}-k_{x}\mathbf{z}) \right) e^ {i(k_{x}x+k_{z}z-\omega t)},
\end{eqnarray}

For the incoming wave, and

\begin{eqnarray}\nonumber
\mathbf{E}_{r} &=& \left(R_{\perp}\mathbf{y}-R_{\parallel}\dfrac{c}{\omega}(k_{z}\mathbf{x}+k_{x}\mathbf{z}) \right) e^{i(k_{x}x-k_{z}z-\omega t)},\\
\nonumber
\mathbf{H}_{r} &=& \left(R_{\parallel}\mathbf{y}+R_{\perp}\dfrac{c}{\omega}(k_{z}\mathbf{x}+k_{x}\mathbf{z}) \right) e^{i(k_{x}x-k_{z}z-\omega t)},
\end{eqnarray}

for the reflected wave, where we have used that $k_{x}^{in}=-k_{x}^{ref}$ with obvious notation. The cartesian unit vectors are represented by $\mathbf{x},\mathbf{y}$ and $\mathbf{z}$ and we define $k_{x}=\frac{\omega}{c}\mathrm{sin}\theta_{i}$ and $k_{z}=\frac{\omega}{c}\mathrm{cos}\theta_{i}$, where $\theta_{i}$ is the angle of incidence. The problem consists in finding the relative amplitudes $A_{\perp},A_{\parallel}$ and $R_{\perp},R_{\parallel}$. Their quotients will define the entries of the reflection matrices \eqref{eq:ReflectionMatrices}.

The second step is to solve Maxwell's equations inside the uniaxial material. Since there is translational invariance along the interface, $k_{x}$ must be conserved so the transmitted wave can have the form:

\begin{eqnarray}
\mathbf{E}_{t} &=& \mathbf{e}e^{i(q z + k_{x}x-\omega t)},\\
\mathbf{H}_{t} &=& \mathbf{h}e^{i(q z + k_{x}x-\omega t)},
\end{eqnarray}

where $q$ is the transverse transmitted momentum to be determined by finding the dispersion relation. From the Maxwell equation $\boldsymbol{\nabla}\times\mathbf{E}=-\dfrac{1}{c}\dfrac{\partial\mathbf{B}}{\partial t}$, with $B_{i}=\mu_{ij}H_{j}$ one can obtain the conditions for vectors $\mathbf{e}(z)$ and $\mathbf{h}(z)$ which are:

\begin{eqnarray}\nonumber
h_{z}=\dfrac{ck_{x}}{\omega\mu_{z}}e_{y}, \\
\label{TE}
h_{x}=-\dfrac{cq}{\omega\mu_{\perp}}e_{y},\\
\nonumber
h_{y}=-\dfrac{c}{\omega\mu_{\perp}}\left(k_{x}e_{z}-qe_{x} \right).
\end{eqnarray}

From the Maxwell equation $\boldsymbol{\nabla}\times\mathbf{H}=\dfrac{1}{c}\dfrac{\partial\mathbf{D}}{\partial t}$ with $D_{i}=\epsilon_{ij}E_{j}$ one obtains:

\begin{eqnarray}\nonumber
e_{z}=-\dfrac{ck_{x}}{\omega\epsilon_{z}}h_{y}, \\
\label{TM}
 e_{x}=\dfrac{cq}{\omega\epsilon_{\perp}}h_{y}, \\
 \nonumber
e_{y}=\dfrac{c}{\omega\epsilon_{\perp}}\left(k_{x}h_{z}-qh_{x} \right).
\end{eqnarray}

From the sets \eqref{TM} and \eqref{TE}, after a few steps one can see that if $e_{y}\neq 0$, $h_{x} \neq 0$ and $h_{y} = 0$ $e_{x}= 0$, $q$ must satisfy:

\begin{equation}
q^2 \equiv q'^2=\left( \dfrac{\omega}{c}\right)^2\mu_{\perp}\epsilon_{\perp}-k_{x}^2\dfrac{\mu_{\perp}}{\mu_{z}}.
\end{equation}

This is just the transverse electric mode inside the material. Then, in this case:

\begin{equation}\label{ey}
e_{y} = -\dfrac{\omega\mu_{\perp}}{cq'}h_{x}.
\end{equation}

In a similar fashion if $e_{y} = 0$, $h_{x} = 0$ and $h_{y} \neq 0$ $e_{x} \neq 0$, $q$ must satisfy:

\begin{equation}
q^2 \equiv q''^2=\left( \dfrac{\omega}{c}\right)^2\mu_{\perp}\epsilon_{\perp}-k_{x}^2\dfrac{\epsilon_{\perp}}{\epsilon_{z}},
\end{equation}

and:
\begin{equation}\label{hy}
h_{y} = \dfrac{\omega\epsilon_{\perp}}{cq''}e_{x},
\end{equation}
which corresponds to the transverse magnetic mode inside the uniaxial crystal.\\
It is now time to impose the boundary conditions. As mentioned above, tangential components of $\bm{E}$ and $\bm{H}$ must be continuous along the interface, thus:

\begin{eqnarray}\label{bounda}
(A_{\parallel}-R_{\parallel})\dfrac{c}{\omega} &=& e_{x},\\
\label{boundb}
(A_{\perp}+R_{\perp}) &=& e_{y},\\
\label{boundc}
(A_{\parallel}+R_{\parallel}) &=& h_{y},\\
\label{boundd}
(R_{\perp}-A_{\perp})\dfrac{c}{\omega} &=& h_{x}.
\end{eqnarray}

The last two equations can be further simplified by using  \eqref{hy} and \eqref{ey} to give:

\begin{eqnarray}\nonumber
(A_{\parallel}+R_{\parallel}) &=& \dfrac{\omega\epsilon_{\perp}}{cq''}e_{x} = \dfrac{\omega\epsilon_{\perp}}{cq''}(A_{\parallel}-R_{\parallel})\dfrac{c}{\omega},\\
\nonumber
(R_{\perp}-A_{\perp})\dfrac{c}{\omega} &=& -\dfrac{cq'}{\omega\mu_{\perp}}e_{y} = -\dfrac{cq'}{\omega\mu_{\perp}}(A_{\perp}-R_{\perp}).
\end{eqnarray}

From these it is clear that the transverse electric and the transverse magnetic mode are decoupled, a well known result from ordinary electromagnetic theory \cite{Born99}. One can now solve for the quotients $R_{\perp}/A_{\perp}$ and $R_{\parallel}/A_{\parallel}$ and find the ordinary reflection coefficients for a uniaxial crystal, which in matrix form define the reflection matrix:
\begin{widetext}
\begin{eqnarray}
\label{eq:anisotropic}
\left(
\begin{array}{c}
  E^{(r)}_{s}  \\
 E^{(r)}_{p}
\end{array} \right) = \left(
\begin{array}{cc}
\dfrac{\mu_{\perp}k_{z}-\sqrt{\dfrac{\omega^2}{c^2}\epsilon_{\perp}\mu_{\perp}-k^2_{\parallel}\dfrac{\mu_{\perp}}{\mu_{z}}}}{\mu_{\perp}k_{z}+\sqrt{\dfrac{\omega^2}{c^2}\epsilon_{\perp}\mu_{\perp}-k^2_{\parallel}\dfrac{\mu_{\perp}}{\mu_{z}}}} & 0\\
  0 & \dfrac{\epsilon_{\perp}k_{z}-\sqrt{\dfrac{\omega^2}{c^2}\epsilon_{\perp}\mu_{\perp}-k^2_{\parallel}\dfrac{\epsilon_{\perp}}{\epsilon_{z}}}}{\epsilon_{\perp}k_{z}+\sqrt{\dfrac{\omega^2}{c^2}\epsilon_{\perp}\mu_{\perp}-k^2_{\parallel}\dfrac{\epsilon_{\perp}}{\epsilon_{z}}}}
\end{array} \right)\left(
\begin{array}{c}
  E^{(i)}_{s}  \\
 E^{(i)}_{p}
\end{array} \right),
\end{eqnarray}
\end{widetext}

where $E^{(r)}_{s,p}$ and $E^{(i)}_{s,p}$ are reflected and transmitted electric field respectively for $s$ and $p$ polarizations.

\subsection*{Fresnel coefficients for uniaxial material plates with a topological magnetoelectric response (axion)}

It is known from earlier works \cite{QHZ08,EMB09} that topological insulators contain, together with a dielectric response, a topological contribution to the magnetoelectric effect originated in an axion type Lagrangean. The inclusion of this coupling into the reflection coefficients was discussed by \cite{Obu05} for the isotropic case. Their conclusion was that the tangential components of $\bm{E}$ and $\bm{H}$ were still conserved, although now $H = \mu^{-1}B + \bar{\alpha} E $ where $\bar{\alpha}$ is the topological magnetoelectric response or axion term and $\alpha$ is the fine structure constant ($\alpha = \frac{e^2}{c\hbar}$). Analogously the inclusion of the new term to the boundary conditions given by \eqref{bounda},\eqref{boundb},\eqref{boundc} and \eqref{boundd} is implemented by modifying the last two equations \eqref{boundc},\eqref{boundd} (which come from the continuity of the tangential component of $H$) to give:

\begin{eqnarray}\label{boundax}
(A_{\parallel}-R_{\parallel})\dfrac{c}{\omega} &=& e_{x},\\
\label{boundbx}
A_{\perp}+R_{\perp} &=& e_{y},\\
\label{boundcx}
A_{\parallel}+R_{\parallel} &=& h_{y} + \bar{\alpha} e_{y},\\
\label{bounddx}
(R_{\perp}-A_{\perp})\dfrac{c}{\omega} &=& h_{x} + \bar{\alpha} e_{x},
\end{eqnarray}

with $h_{x}$ given by \eqref{ey} and $h_{y}$ given by \eqref{hy}. It is now a matter of algebra to elucidate the reflection coefficients as in the last section. Inevitably, the relations become more messy. The reflection coefficients can be written in matrix form as:

\vspace{-0.7cm}
\begin{widetext}
\begin{eqnarray}
%\label{eq:anisotropicandTI}
\left(
\begin{array}{c}
  E^{(r)}_{s}  \\
 E^{(r)}_{p}
\end{array} \right) =\dfrac{1}{\Delta_{an}}\left(
\begin{array}{cc}
\left(\mu_{\perp}k_{z}-q'\right) \left(\epsilon_{\perp}k_{z}+q'' \right)-q''k_{z}\mu_{\perp}\bar{\alpha}^2  & 2\bar{\alpha}q''\mu_{\perp}k_{z}\\
  2\bar{\alpha}q''\mu_{\perp}k_{z} & \left( \epsilon_{\perp}k_{z}-q''\right) \left(\mu_{\perp}k_{z}+q' \right)+q''k_{z}\mu_{\perp}\bar{\alpha}^2
\end{array} \right)\left(
\begin{array}{c}
  E^{(i)}_{s}  \\
 E^{(i)}_{p}
\end{array} \right),
\end{eqnarray}
\end{widetext}
where $\Delta_{an}= \left( \mu_{\perp}k_{z}+q'\right) \left(\epsilon_{\perp}k_{z}+q'' \right)+q''k_{z}\mu_{\perp}\bar{\alpha}^2 $, $k^2_{z} = \dfrac{\omega^2}{c^2}-k^{2}_{\parallel}$, $q'^{2} = \dfrac{\omega^2}{c^2}\epsilon_{\perp}\mu_{\perp}-k^2_{\parallel}\dfrac{\mu_{\perp}}{\mu_{z}}$, $q''^{2} = \dfrac{\omega^2}{c^2}\epsilon_{\perp}\mu_{\perp}-k^2_{\parallel}\dfrac{\epsilon_{\perp}}{\epsilon_{z}}$ and $\bar{\alpha} = \alpha\theta/\pi$.\\
To compute the Casimir energy, one should define these in the imaginary frequency axis and turn all the frequencies to $\omega = i\xi$. Note as well that these reduce to the ordinary anisotropic coefficients presented in the last section when $\bar{\alpha}=0$. They also reduce to the isotropic reflection coefficients \eqref{eq:ReflectionMatricesTIcomplete} when $\mu$ and $\epsilon$ are isotropic.
%\bibliography{Casimir.bib}
\newcommand{\npb}{Nucl. Phys. B}\newcommand{\adv}{Adv.
  Phys.}\newcommand{\epl}{Europhys. Lett.}

\end{document}